\documentclass{medphyspaper}
\usepackage{graphicx}
\usepackage{caption}
\usepackage{subcaption}
\usepackage{float}
\usepackage{wrapfig}
\usepackage[export]{adjustbox}
\usepackage{adjustbox}
\usepackage{xcolor}
\usepackage{multirow}
\usepackage{booktabs}
\usepackage{longtable}
\usepackage{array}
\usepackage{arydshln}
\usepackage{afterpage}
\usepackage{csquotes}
\usepackage{siunitx}
\usepackage{scalefnt}

\usepackage{comment}
\usepackage{gensymb}
\usepackage{pgfplots}
  \pgfplotsset{compat=newest}
  \usetikzlibrary{plotmarks}
  \usetikzlibrary{arrows.meta}
  \usepgfplotslibrary{patchplots}
  \usepackage{grffile}
  \usetikzlibrary{external}
  \usetikzlibrary{pgfplots.statistics, pgfplots.colorbrewer} 

\usepackage{pgfplotstable}
\usepackage{filecontents}
\usepackage{cleveref}

\usepackage{subfiles}

\newcommand{\sqDev}{\textit{$f_{sq \ deviation}$}}
\newcommand{\sqOve}{\textit{$f_{sq \ over \ dosage}$}}
\newcommand{\mVar}{\textit{$f_{mean \ Variance}$}}
\newcommand{\mVarc}{\textit{$c_{mean \ Variance}$}}
\newcommand{\maxDVH}{\textit{$f_{max DVH}$}}
\newcommand{\meanD}{\textit{$f_{mean}$}}
\newcommand{\cMaxMean}{\textit{$c_{max \ mean \ dose}$}}

\usepackage{xparse}
\usetikzlibrary{external}
\newif\iftikzforceexternal
\newcommand{\subfolder}{.}
\NewDocumentCommand{\includetikz}{O{.tex} O{\subfolder/} O{\subfolder/tikz_ext/} m}{%
	\iftikzforceexternal%
		\includegraphics{#3#4.pdf}%	
	\else%
		\tikzsetnextfilename{#3#4}%
		\input{#2#4#1}%
	\fi%
}
\tikzset{%
	every picture/.append style={%
			line join=round,%
	}%
}%
\tikzforceexternaltrue

\title{Scenario-free robust optimization algorithm for IMRT and IMPT treatment planning} 
\runningtitle{Scenario-free robust planning}

\author[DKFZ,HIRO, HDPHYS]{Remo~Cristoforetti} %The star means shared contribution
\ead{remo.cristoforetti@dkfz-heidelberg.de}

\address[DKFZ]{Department of Medical Physics in Radiation Oncology, German Cancer Research Center -- DKFZ, Im Neuenheimer Feld 280, 69120 Heidelberg, Germany}
\address[HIRO]{Heidelberg Institute for Radiation Oncology -- HIRO, Im Neuenheimer Feld 280, 69120 Heidelberg, Germany}

\author[DKFZ, HIRO, HDPHYS]{Jennifer~Josephine~Hardt}
\address[HDPHYS]{Faculty of Physics and Astronomy, Heidelberg University, Heidelberg, Germany}

\author[DKFZ,HIRO]{Niklas Wahl}
\ead{n.wahl@dkfz-heidelberg.de}

\version{1}

\begin{document}

\maketitle

\begin{abstract}
{\textbf{Background:}}
Robust treatment planning algorithms for Intensity Modulated Proton Therapy (IMPT) and Intensity Modulated Radiation Therapy (IMRT) allow for uncertainty reduction in the delivered dose distributions through explicit inclusion of error scenarios. Due to the curse of dimensionality, application of such algorithms can easily become computationally prohibitive.

{\textbf{Purpose:}}
This work proposes a scenario-free probabilistic robust optimization algorithm that overcomes both the runtime and memory limitations typical of traditional robustness algorithms.

{\textbf{Methods:}}
The scenario-free approach minimizes cost-functions evaluated on expected-dose distributions and total variance. Calculation of these quantities relies on precomputed expected-dose-influence and total-variance-influence matrices, such that no scenarios need to be stored for optimization. The algorithm is developed within matRad \cite{matRad, Wieser2017} and tested in several optimization configurations for photon and proton irradiation plans. A traditional robust optimization algorithm and a margin-based approach are used as a reference to benchmark the performances of the scenario-free algorithm in terms of plan quality, robustness and computational workload. 

{\textbf{Results:}}
The implemented scenario-free approach achieves plan quality compatible with traditional robust optimization algorithms and it reduces the distribution of standard deviation within selected structures when variance reduction objectives are defined. Avoiding the storage of individual scenario information allows for the solution of treatment plan optimization problems including an arbitrary number of error scenarios. The observed computational time required for optimization is compatible with a nominal, non-robust algorithm and significantly lower compared to the traditional robust approach. Estimated gains in relative runtime range from approximately \num{5} to \num{600} times with respect to the traditional approach.

{\textbf{Conclusion:}}
This work demonstrates how the scenario-free optimization algorithm can achieve the required dose and robustness specifications under multiple different optimization conditions. The measured runtime and memory footprint are independent on the number of included error scenarios and compatible with those of non-robust margin-based optimization algorithms. These properties make the scenario-free approach suitable for the optimization of robust plans involving an high number of error scenarios and CT phases as 4D robust optimization.    
\end{abstract}

%\mainmatter

% Introduction section
\newpage
\section{Introduction}
The effective handling of uncertainty poses a fundamental challenge for external beam radiotherapy treatment planning.\cite{IAEA2016} Nevertheless, modeling the multiple sources of uncertainty and implementing feasible robust optimization strategies is beneficial to preserve plan quality  in case of errors manifesting during treatment.\cite{Unkelbach2018}

Different treatment modalities, tumor sites and patient anatomy require different approaches to handling uncertainties during computational treatment planning when they can not be mitigated in preparation and setup.\cite{vanHerk2004,Unkelbach2007,Lomax2008,Lomax2008a,Fredriksson2012,Unkelbach2018} When focusing on treatment plan optimization, the gold standard approach to ensure target coverage relies on the definition of planning margins. This approach has proven reliable under the assumption that the static dose cloud approximation holds and is particularly effective for intensity modulated radiation therapy (IMRT).\cite{Stoll2016,vanHerk2000}

The static dose cloud approximation assumes that rigid shifts of the dose distribution well approximate the impact of main sources of uncertainties such as set up errors or organ motion. It is generally only valid for regions of homogeneous density and target sites of reproducible positioning. \cite{Fredriksson2016} Such assumption cannot generally be extended to the case of protons and heavy charged particle irradiation, given the high sensitivity to range uncertainty.\cite{Lomax2008,Lomax2008a}

For cases in which the static dose cloud approximation breaks down, robust optimization algorithms were developed to explicitly include a representation of the uncertainty model into the optimization problem.\cite{Fredriksson2012,Unkelbach2018} The scenario-based representation is a generalization of the margin-concept for the non-static case\cite{Fredriksson2016, Korevaar2019} and thus bypasses the definition of planning margins with the optimization focusing on the Clinical Target Volume (CTV).

Robust optimization is particularly beneficial in highly non-static treatment configurations such as lung cancer irradiation. \cite{Ge2019,Liu2016} Additional uncertainty introduced by the breathing motion can be accounted for by including 4D-CT datasets, thus performing a 4D-robust optimization. Use of 4D-robust optimization can improve target coverage and OAR sparing as opposed to non-robust optimization based on margin definition. \cite{Wang2023}

Different mathematical formulations of the optimization problem can be exploited to achieve robustness goals,\cite{Fredriksson2012,Bangert2013,Wahl2018a,Taasti2020} however, the computational implementation of such algorithms generally relies on, and is burdened by, the estimation of multiple error scenario distributions.\cite{Fu2023} 
A statistically stable representation of the uncertainty model requires a sufficiently large amount of error scenarios to be sampled from the underlying probability distribution.\cite{Wahl2017, Wahl2018a} Therefore, such algorithms always imply a trade-off between efficacy and feasibility.

Typical examples of robust optimization algorithms are the \textit{minMax} \cite{Fredriksson2011} and the \textit{stochastic},\cite{Unkelbach2009} sometimes also called \textit{expected value}, approaches.\cite{Fredriksson2012,Unkelbach2018} These algorithms have proven to be more effective with respect to margin based approaches in achieving dose conformity and reducing the impact of uncertainties on the optimized distributions \cite{Liu2013,Yagihashi2024}. At the same time, they bear intrinsic applicability limitations due to high memory and computational time demand.

Optimization time constraints are of particular concern, especially when 4D-robust optimization is applied. \cite{Wolf2020}

The present work explores the capabilities of an alternative, \enquote{scenario-free} robust optimization algorithm. With this approach, the optimization problem's design does not rely anymore on the explicit evaluation of individual scenario distributions. The approach is derived from previous works on probabilistic dose calculation and optimization\cite{Bangert2013,Wahl2017,Wahl2018a} and the expected value approach. It relies on precomputation of expected dose and variance influence terms and thus not requires the storage and computation of multiple scenarios during optimization. Dosimetric objectives are therein defined on the expectation value of the dose while variance-reduction objectives are applied to each structure to minimize the mean variance value.

The key advantage of our approach is that the dimensionality of the pre-computed expected dose and variance influence objects is independent of the number of scenarios (in the case scenarios are used for their construction). This way, the optimization becomes scenario-free and the approach addresses simultaneously the memory limitation issue and the increased computational time demand of iterating through scenarios during each iteration in optimization.

The algorithm was tested in several treatment configurations. Results were initially collected for the simplified setup of a cubic, homogeneous phantom and extended to clinically realistic cases of lung cancer patients. In both cases, 4D datasets were included to highlight the feasibility of applying the algorithm to otherwise unfeasible optimization configurations. Comparison of plan quality and robustness metrics was performed with respect to margin-based and traditional robust optimization algorithms.

% Material and methods
\section{Materials and Methods}
\subsection{Robust optimization}
In its general formulation, the robust treatment planning optimization problem aims at finding optimal beamlet intensities $\boldsymbol{x}^*$ by solving
\begin{equation}
\begin{aligned}
    \boldsymbol{x}^* &= \operatornamewithlimits{arg\, min}_{\boldsymbol{x}}\ \mathcal{R}_\mathcal{U}\left[F(\boldsymbol{x})\right]\\
    \mathrm{s.t.}\qquad \boldsymbol{x} &> 0\\
    c_n(\boldsymbol{x}) &\leq 0
\end{aligned}
\label{eq:opt-prob}
\end{equation}
where $F$ denotes the global cost-function to be minimized depending on the beamlet intensity vector $\boldsymbol{x}$. $\mathcal{R}_\mathcal{U}$ denotes a robustness operator acting on $F$ depending on an uncertainty model $\mathcal{U}$, and the problem can be subjected to by multiple constraint functions $c_n$.

$F$ usually represents a multi-criteria decision function, e.\,g., via the weighted-sum scalarization $F = \sum_{i} p_i f_i$ of multiple objectives $f_i$ weighted with $p_i$. Most if not all of these terms will depend on the dose distribution $\boldsymbol{d}({x}) = \boldsymbol{D}\boldsymbol{x}$, computed as a linear transformation of $\boldsymbol{x}$ with a precomputed \emph{dose-influence matrix} $\boldsymbol{D}$.

As dose is the main quantity affected by the uncertainty model, applying $R_\mathcal{U}$ usually observes possible variation in $d$ due to the uncertainties $\mathcal{U}$. A common way of doing so consists in sampling an arbitrary amount of \textit{error scenarios} from a probability distribution describing the uncertainty model $\mathcal{U}$.
The robust operator $\mathcal{R}$ then can take many forms, with the most prominent being \emph{minimax}, \emph{Conditional Value at Risk}, or \emph{expected value}. For the approach presented in this manuscript, let us consider the latter expected value, often also named \emph{probabilistic}, approach of
\begin{equation}
    \mathcal{R}_\mathcal{U}\left[F(\boldsymbol{d}({x}))\right] = \mathbb{E}\left[F(\boldsymbol{d}({x}))\right] \approx \sum_{s} \pi_s \left[F(\boldsymbol{d}_{s\in\mathcal{S}}({x}))\right]\ .
    \label{eq:rob-opt-prob}
\end{equation}

\Cref{eq:rob-opt-prob} approximates the expectation value operator $\mathbb{E}$ by evaluation of a set of error scenarios $\mathcal{S}$ with importance weights $\pi_s$. Note that, in general, evaluation of each dose scenario $\boldsymbol{d}_s = \boldsymbol{D}_s\boldsymbol{x}$ entails storing a scenario dose-influence matrix $\boldsymbol{D}_s$.
\\
\subsection{Scenario-free robust optimization}
\subsubsection{Reformulation of the probabilistic optimization problem}
The developed scenario-free probabilistic approach is based on previous work by \citet{Bangert2013}, which used a probabilistic dose calculation algorithm to find a fully analytical approximation of the expectation value operator in \cref{eq:rob-opt-prob} using a precomputed expected dose influence matrix $\mathbb{E}[\boldsymbol{D}]$ and total variance influence matrices $\Omega$.

To do so, \citet{Bangert2013} started from the special case of the penalized least-squares objective function

\begin{equation}
    F(\boldsymbol{d}) = (\boldsymbol{d} - \boldsymbol{d^{ref}})^{T} P (\boldsymbol{d} - \boldsymbol{d^{ref}}) %= p \sum_{i} (d_{i} - d^{ref}_{i})^{2}
    \label{eq:squared-deviation}
\end{equation}
and its expectation value
\begin{equation}
    \mathbb{E}[F] = \operatorname{tr}(\boldsymbol{P}\boldsymbol{\Sigma}_{\boldsymbol{d}}) + (\mathbb{E}[\boldsymbol{d}] - \boldsymbol{d^{ref}})^{T} \boldsymbol{P} (\mathbb{E}[\boldsymbol{d}] - \boldsymbol{d^{ref}})\ .
    \label{eq:exp-value-f}
\end{equation}
In \cref{eq:squared-deviation,eq:exp-value-f}, $\boldsymbol{d}^{ref}$ is the prescribed dose distribution and $P = \operatorname{diag}(p_1, p_2, \dots, p_N)$ is a diagonal matrix assigning penalty factors $p_i$. \Cref{eq:exp-value-f} is composed of two terms; a \enquote{total variance} term $\operatorname{tr}(\boldsymbol{P}\boldsymbol{\Sigma}_{\boldsymbol{d}})$ representing the penalty-weighted sum of variance ($\boldsymbol{\Sigma}_{\boldsymbol{d}}$ is the dose's covariance matrix), and the penalized quadratic difference of the expectation value to the prescription.

Evaluation of both terms can now be facilitated by aforementioned total variance influence $\Omega$ and expected dose influence matrix $\mathbb{E}[\boldsymbol{D}]$. While this is trivial for the second term using
\begin{equation}
    \mathbb{E}[\boldsymbol{d}] = \mathbb{E}[\boldsymbol{D}]\boldsymbol{x}\label{eq:expected-dose-influence}
\end{equation}
similarly to nominal dose evaluation, rewriting the variance term requires a closer look using multiple properties of the matrix trace:

\begin{equation}
\begin{aligned}
    \operatorname{tr}(\boldsymbol{P}\boldsymbol{\Sigma}_{\boldsymbol{d}})
    &= \operatorname{tr}\left(\boldsymbol{P}(\mathbb{E}[\boldsymbol{d}\boldsymbol{d}^T] - \mathbb{E}[\boldsymbol{d}]\mathbb{E}[\boldsymbol{d}]^T)\right)\\
    &= \operatorname{tr}\left(\boldsymbol{P}\left(\mathbb{E}[\boldsymbol{D}\boldsymbol{x}\boldsymbol{x}^T\boldsymbol{D}^T] - \mathbb{E}[\boldsymbol{D}]\boldsymbol{x}\boldsymbol{x}^T\mathbb{E}[\boldsymbol{D}]^T\right)\right)\\
    &= \operatorname{tr}\left(\boldsymbol{x}\boldsymbol{x}^T\mathbb{E}[\boldsymbol{D}^T\boldsymbol{P}\boldsymbol{D}] - \boldsymbol{x}\boldsymbol{x}^T\mathbb{E}[\boldsymbol{D}]^T\boldsymbol{P}\mathbb{E}[\boldsymbol{D}]\right)\\
    &= \operatorname{tr}\left(\boldsymbol{x}\boldsymbol{x}^T\mathbb{E}[\boldsymbol{D}^T\boldsymbol{P}\boldsymbol{D}]\right) - \operatorname{tr}\left(\boldsymbol{x}\boldsymbol{x}^T\mathbb{E}[\boldsymbol{D}]^T\boldsymbol{P}\mathbb{E}[\boldsymbol{D}]\right)\\
    &= \boldsymbol{x}^T \mathbb{E}[\boldsymbol{D}^T\boldsymbol{P}\boldsymbol{D}] \boldsymbol{x} - \boldsymbol{x}^T \mathbb{E}[\boldsymbol{D}]^T\boldsymbol{P}\mathbb{E}[\boldsymbol{D}] \boldsymbol{x}\\
    &= \boldsymbol{x}^T\left(\mathbb{E}[\boldsymbol{D}^T\boldsymbol{P}\boldsymbol{D}] - \mathbb{E}[\boldsymbol{D}]^T\boldsymbol{P}\mathbb{E}[\boldsymbol{D}]\right)\boldsymbol{x}\\
    &= \boldsymbol{x}^T\boldsymbol{\Omega}\boldsymbol{x}\ . %\boldsymbol{P} \big( \mathbb{E}[\boldsymbol{d}^{2}] - \mathbb{E}[\boldsymbol{d}]^{2}\big)aligned}
\end{aligned}
\label{eq:trace_term}
\end{equation}

While pre-computing $\boldsymbol{\Omega}$ according to \cref{eq:trace_term} would factor in the penalty factors for each voxel into the pre-computation, planning reality usually defines penalty factors per volume of interest (VOI). This, in turn, allows us to separate $\boldsymbol{\Omega}$ VOI-wise summands of the matrix $\boldsymbol{\Omega}$ without including the penalties:
\begin{equation}
    \Omega = \sum_v p_v\Omega_v = p_v\left(\mathbb{E}[\boldsymbol{D}_v^T\boldsymbol{D}_v] - \mathbb{E}[\boldsymbol{D}_v]^T\mathbb{E}[\boldsymbol{D}_v]\right)
\end{equation}

The heavy matrix products are then carried out including only the voxels within the respective structure. Once normalized by the structure volume, the quadratic form $\boldsymbol{x^{T}}\boldsymbol{\Omega}\boldsymbol{x}$ represents a \emph{total variance} term associated to the structure and quantifies the uncertainty of the dose distribution.
$\boldsymbol{\Omega}$ is a square, positive semi-definite matrix. The largest benefit of introducing $\Omega$ is that its size is independent of the number of voxels for optimization, while the most substantial drawback is the loss of voxel-specific information about the variance.

Now, \cref{eq:exp-value-f} is of course only strictly valid when the squared deviation cost function (\cref{eq:squared-deviation}) is applied. However, such a formulation allows to interpret the two terms appearing in \cref{eq:exp-value-f} as two independent cost-functions. The first one encodes a variance-reduction objective, the second one encodes the dosimetric objective. Following such interpretation, a new optimization problem can be defined by extending the application of such cost-functions to any dosimetric objective. The scenario-free optimization approach consists thus in applying the desired cost-function to the expectation value of the dose distribution (estimated via \cref{eq:expected-dose-influence}) and in adding independently the variance-reduction term.

\subsubsection{Precalculation of the probabilistic quantities}\label{subsection:MM_calc_prob_quantities}
Note that the term "scenario-free" only applies to optimization, while during dose calculation explicit calculation of each error scenario still occurs. However, accumulation of the probabilistic quantities $\boldsymbol{\Omega}$ and $\mathbb{E}[\boldsymbol{D}]$ can be carried out to avoid retention of individual scenario information in memory. This way, the pure optimization time still remains independent of the number of considered scenarios.

A fundamental role in the definition of the probabilistic quantities is played by the procedure applied to sample the probability distribution modeling the uncertainty source. This work explores both a worst-case and a random sampling approach. For the first case, \num{26} shift error scenarios were sampled. \num{2} additional over- and under- shoot range errors and the nominal scenario were added for a total amount of \num{29} scenarios included \cite{Korevaar2019}. For the random sampling approach instead, the number of scenarios can be defined by the user and sample size analysis was performed to showcase the impact of such choice on plan robustness.

In case a 4DCT dataset is included, two different approaches can be followed for handling the different CT-phases. Error scenarios can indeed be accumulated either within each CT-phase or considering the whole pool of error and CT scenarios combined. Both methods were implemented and compared.

\subsection{Implementation and evaluation}
The algorithm was developed and tested within \textit{matRad}\cite{matRad, Wieser2017}. The implemented scenario-free approach was benchmarked against the traditional expected value optimization method described by \cref{eq:rob-opt-prob} including multiple 3D and 4D robust optimization setups. In the following, such algorithm will be referred to as \textit{stochastic}. For the 4D optimization strategies, an additional non robust, margin-based plan was also developed. In this case, to partially preserve target coverage in presence of uncertainty, a suitable margin was applied to the internal target volume (ITV) to obtain the planning target volume (PTV).
\\
\\
Setup error scenarios were obtained by a rigid shift of the isocenter position in all three spatial directions. The translation amplitudes were sampled from a normal probability distribution with a set value of $2.25 \ mm$ standard deviation. Range errors were instead modeled as an absolute and relative deviation of the water-equivalent voxel depths. Deviations were sampled again from normal probability distributions with standard deviations of $1 \ mm$ and $3.5 \%$ respectively. For the worst case sampling, the same probability distributions were used and scenarios were collected at $\pm 2 \ \sigma$ both for setup and range errors.
\\
\\
For each optimized plan, robustness analysis was performed evaluating the scenario-specific dose distribution with the optimized set of fluence weights. Robustness was quantified through Standard-Deviation-Volume-Histograms (SDVH). An SDVH relates standard deviation (SD) to percentage volume and is used to graphically represent the impact of uncertainty within a given structure. Additionally, Dose-Volume-Histogram (DVH) bands, dose and SD distributions are reported for a subset of cases.

The computational efficiency in solving the optimization problem was estimated in terms of time per iteration (TPI), i.\,e., the average time required to compute one step of the iterative optimization problem.
Taking the nominal margin-based optimization as a reference, performances among different algorithms can be quantified in terms of relative TPI (rTPI).

\subsection{Treatment Plans}\label{subsection:Treatment_plans}
Four geometrical models were included to showcase different features of the developed algorithm. These included respectively a simple homogeneous box phantom, both in a 3D and 4D setup, and two lung cancer patient 4DCT datasets. The homogeneous box phantom includes a central box-shaped target the HU values of which have been increased to simulate a difference in density between target and OARs.
The 4D dataset for the box-shaped phantom was obtained by artificially simulating an horizontal organ motion of the 3D box-phantom through multiple CT-phases.

The patient datasets were instead selected from the Cancer Imaging Archive \cite{data4DLung2016}. Geometrical specifications of the investigated phantoms and patinet cases are summarized in \cref{tab:geom-phantoms}. Plans were optimized for both photon and proton irradiation. For the proton cases, a constant $RBE$ factor of \num{1.1} was applied.
\begin{table}[h]
    \centering
    \caption{Properties of the different geometrical setup used. \#CTs refers to the number of CT-phases available, while (p$^+$) and ($\gamma$) refer respectively to proton and photon irradiation.}\label{tab:geom-phantoms}
    \footnotesize
    \begin{tabular}{ccccc}
    \toprule
         Phantom & \#CTs & Field angles (p$^+$) & Field angles ($\gamma$) & ITV-margin\\
         \midrule
         BOX & 1 & 0$\degree$, 90$\degree$ & - &  -\\
 %
         %\midrule
         4D-BOX  & 10 & 0$\degree$, 90$\degree$ & $0\degree$, $51 \degree$,$102 \degree$,$154 \degree$,$205 \degree$,$257 \degree$, $308 \degree$ & $4 \ mm$\\
         %\midrule
         Patient 1  & 10 &  $45 \degree$, $90 \degree$, $135 \degree$ & $0\degree$, $51 \degree$,$102 \degree$,$154 \degree$,$205 \degree$,$257 \degree$, $308 \degree$ & $4 \ mm$\\
         %\midrule
         Patient 2 & 10 & $0 \degree$, $270 \degree$, $315 \degree$ & $0\degree$, $51 \degree$,$102 \degree$,$154 \degree$,$205 \degree$,$257 \degree$, $308 \degree$ & $8 \ mm$\\
        \bottomrule
    \end{tabular}
    
\end{table}

The  applied cost functions and reference dose prescriptions varied according to the optimization setup implemented. The traditional probabilistic algorithm and the margin-based approach always apply the cost-functions to the individual scenario distributions. For all the optimized plans, the scenario-free approach consistently applies the same dosimetric objectives to the expected dose distribution. Robustness is achieved via additional variance reduction objectives or constraints, the relative importance of which is tuned accordingly. An overview of the optimization configurations used is reported in \cref{tab:cost-fun}.

\begin{table}[H]
    \centering
    \caption{Summary of the optimized plans for all phantoms, including the number of scenarios used for optimization, and the optimization algorithm applied: scenario-free (s-f), stochastic (s), and margin-based (m-b). According to the specific phantom, the cost-functions are applied to different structures referred as: Target (T), OAR (O), Lungs (L), Spinal Cord (SC), and Body contour (B). The applied cost functions are reported as defined in \cite{matRad}, with the additional mean variance objective and constraints (\mVar, \mVarc) and a constraint set for the maximum value of mean dose (\cMaxMean).}\label{tab:cost-fun}
    \footnotesize
    \begin{tabular}{ccccl}
        \toprule
        \textbf{Phantom} & \multicolumn{1}{c}{\textbf{Description}} & \textbf{Algorithms} & \textbf{Scenarios} & \multicolumn{1}{c}{\textbf{Cost-Functions}} \\
        \midrule

        \multirow{5}{*}{BOX} 
        & \multirow{2}{*}{Algorithm validation} & \multirow{2}{*}{s, s-f} & \multirow{2}{*}{9} & (T) \sqDev, \mVar\\
        & & & & (O) \sqDev, \mVar\\
        & Variance constrained  & \multirow{2}{*}{s-f} & \multirow{2}{*}{9} & (T) \sqDev, \mVarc \\
        & optimization & & & (O) \sqDev, \mVar\\[1ex]
        & Worst-case vs Random & \multirow{2}{*}{s, s-f} & \multirow{2}{*}{29} & \\
        & sampling & & & (T) \sqDev, \mVar,\\
        & \multirow{2}{*}{Sample size analysis} & \multirow{2}{*}{s-f} & 5, 29, 50, &  (O) \sqOve, \mVar\\
        &  & & 100 &  \\
        \midrule

        \multirow{4}{*}{4D-BOX} 
        & \multirow{2}{*}{4D Scenario accumulation}& \multirow{2}{*}{s-f} & \multirow{2}{*}{30} & (T) \sqDev, \mVar, \\
        & & & & (O) \sqOve, \mVar \\[2ex]
        & \multirow{2}{*}{Plan comparison} & \multirow{2}{*}{s, s-f, m-b} & \multirow{2}{*}{30} & (T) \sqDev, \mVar, \\
        & & & & (O)  \sqOve, \mVar \\

        \midrule

        \multirow{2}{*}{P1} 
       & \multirow{2}{*}{Plan comparison} & \multirow{2}{*}{s, s-f, m-b} & \multirow{2}{*}{30} & (T) \sqDev, \mVar, \\
        & && & (O)  \sqOve \\

        \midrule

        \multirow{5}{*}{P2} 
        & \multirow{5}{*}{Plan comparison} & \multirow{5}{*}{s-f, m-b} & \multirow{5}{*}{100} & (T)\ \ \ \sqDev, \mVar \\
        & & & & \multirow{2}{*}{(L)}\ \ \ \maxDVH, \meanD,\\
        & & & & \ \ \ \ \ \ \ \mVar, \cMaxMean  \\
        & & & &(SC) \meanD, \mVar \\
        & & & & (B)\ \ \ \sqOve \\
        \bottomrule
    \end{tabular}
\end{table}

% results
\section{Results}
    \subsection{Box Phantom}
    \subsubsection{Algorithm validation on the box-phantom}
The first plan was optimized for validation purposes on the box phantom. It uses two proton beams, and was optimized both with the scenario-free and the traditional stochastic robust algorithm. The applied cost-functions are pure penalized least-squares objectives of the form expressed by \cref{eq:exp-value-f}. Additionally, a nominal non-robust plan was also optimized for comparison using \cref{eq:squared-deviation}. In order to enhance the impact of uncertainty, no additional planning margin was added for this plan to the target structure to obtain the PTV.

Robustness analysis was performed on the optimized plans over a pool of \num{100} setup and range error scenarios randomly sampled from the same probability distribution used to accumulate the probabilistic quantities.

\begin{figure}[h]
        \centering
        \resizebox{\textwidth}{!}{
            \includegraphics{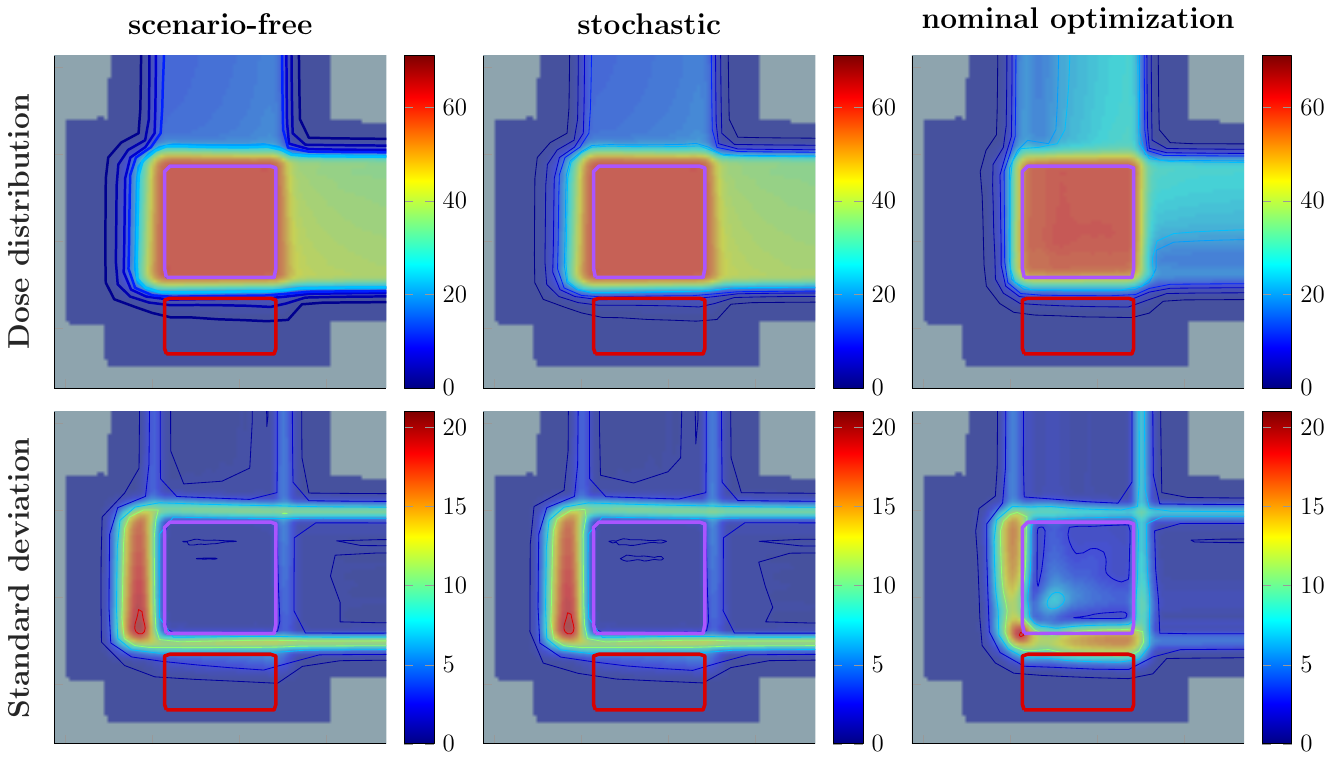}
        }
        \caption{Expected dose (top) and corresponding SD (bottom) obtained with the scenario-free (left), traditional stochastic (middle) and nominal (right) optimization approaches. The target structure and the OAR are contoured in purple and red respectively. All the colorbar values are reported in \si{\gray}.}
        \label{fig:box_dose_dist}
    \end{figure}

The obtained dose and standard deviation distribution for the three algorithms are depicted in \cref{fig:box_dose_dist} for a single isocentric slice of the phantom The central cubic target and a distal OAR are contoured in purple and red respectively.

Both the robust optimization algorithms achieved a low and uniform distribution of uncertainty within the target structure. In this case of pure penalized least-squares objective functions, both optimization problems are equivalent: A gamma analysis (\SI{3}{mm}, $3\%$) confirmed this with a 100\% pass-rate for both dose and SD distributions.
The nominal reference plan showed higher values of SD within the target and toward the distal range for both fields.

\Cref{fig:box_target_DVH_SDVH_a} reports the DVHs and SDVHs analysis performed on the target structure for both robust optimization strategies and the nominal plan. This confirms the high target dose uncertainty observed in the nominal reference plan, while the equivalent stochastic and scenario-free approaches show reduced, nearly equal expected DVH and confidence bands. The evolution of the objective functions through the iterations of the algorithm is depicted in \cref{fig:box_target_DVH_SDVH_b}, further highlighting the equivalence of the robust optimization algorithms.
\begin{figure}[H]
    \begin{subfigure}[b]{0.49\textwidth}
            \resizebox{1.03\textwidth}{!}{
                \includegraphics{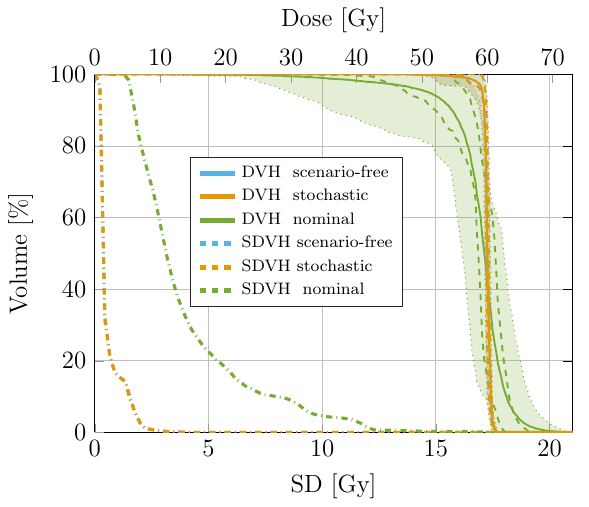}
            }
        \caption{Target DVHs}
        \label{fig:box_target_DVH_SDVH_a}
    \end{subfigure}
    \hfill
    \begin{subfigure}[b]{0.49\textwidth}
        \raisebox{0.3cm}{
            \resizebox{\textwidth}{!}{
                \includegraphics{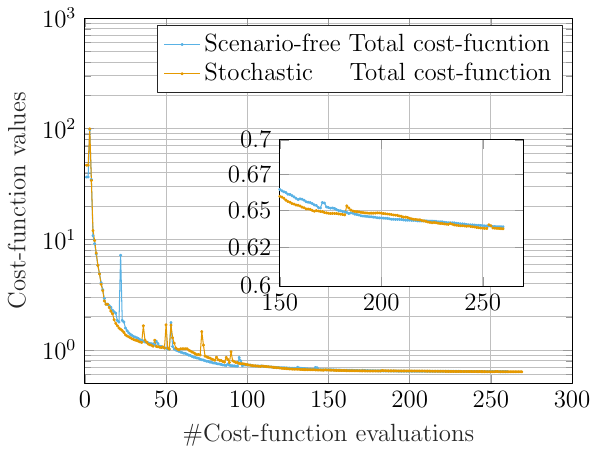}
            }
        }
        \caption{Cost-function values}
        \label{fig:box_target_DVH_SDVH_b}
    \end{subfigure}
    \caption{Dosimetric quality and cost-function evolution for the validation of the scenario-free algorithm compared to the conventional stochastic implementation. (a) DVHs and SDVHs for the target structure and the two applied algorithms. The solid line represents the DVH computed for the expected dose distribution while the dashed and dotted lines correspond to the $25$-$75$ and the $5$-$95$ percentiles of single-scenario DVHs distributions. (b) Values of the cost-functions for each performed iteration for the two robust algorithms.}
    \label{fig:box_target_DVH_SDVH}
\end{figure}

Although the absolute optimization time strongly depends on the specific implementation and available hardware, the rTPI for the two robust plans can still be compared. For this plan, the rTPIs for the scenario-free and stochastic approach were respectively of \num{1.7} and \num{8.3}.

\subsubsection{Constrained scenario-free optimization}
The second experiment on the box phantom validates constrained variance minimization. Two plans were optimized setting mean variance constraints for the target structure with upper thresholds of $\bar{\sigma}^{2}_{a} = \SI{2.9e-2}{\gray\squared}$ and $\bar{\sigma}^{2}_{b} = \SI{2.9e-3}{\gray\squared}$, respectively.
\Cref{fig:box_target_DVH_SDVH} compares both variance constraints used in scenario-free optimization.
\begin{figure}[H]
    \begin{subfigure}[b]{0.5\textwidth}
        \hspace*{-0.5cm}
        \resizebox{\textwidth}{!}{
            \includegraphics{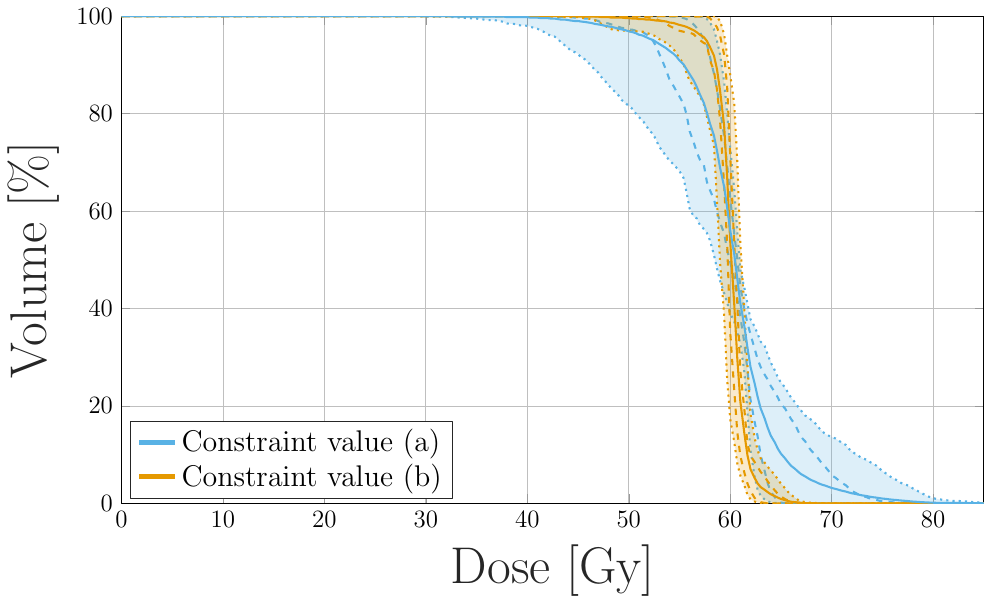}
        }
        \caption{Target DVHs}
        \label{fig:box_target_DVH_SDVH_a}
    \end{subfigure}
    \begin{subfigure}[b]{0.5\textwidth}
        \resizebox{\textwidth}{!}{
            \includegraphics{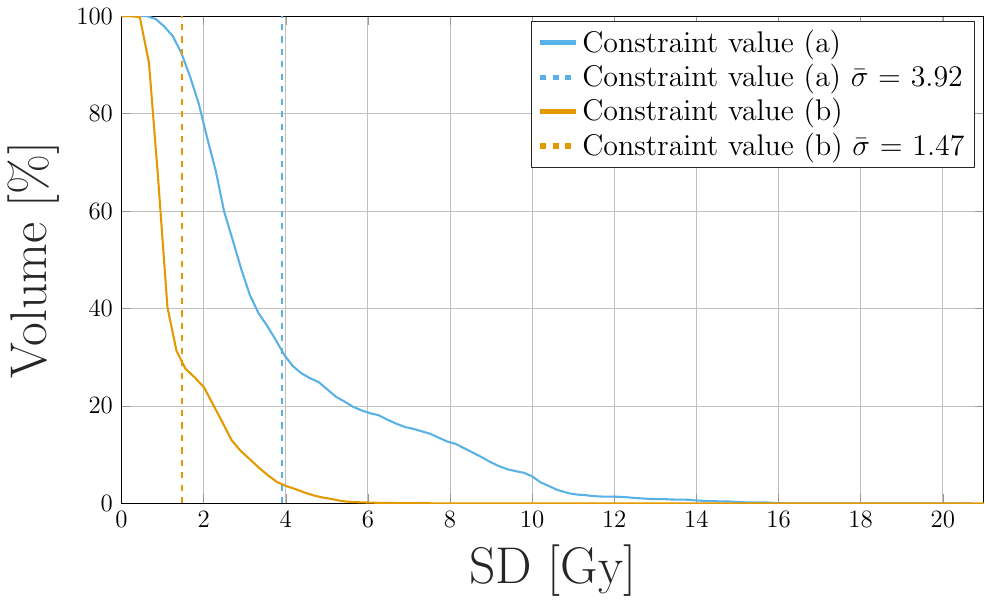}
        }
        \caption{Cost-function values}
        \label{fig:box_target_DVH_SDVH_b}
    \end{subfigure}
    \caption{DVH (a) and SDVH (b) comparison for the target structure when two different constraint values are applied to the \enquote{mean variance} term in the cost-function. The dotted vertical lines in the SDVH plot represent the mean standard deviation, i.e., the average value of the SD distribution.}
    \label{fig:box_target_DVH_SDVH}
\end{figure}

As expected, posing a tight constraint on the mean variance results in a shift toward lower values of average standard deviation. At the same time, the distribution of DVHs also cluster around the dose prescription of \SI{60}{\gray}.

\subsubsection{Algorithm extension and sampling technique analysis}
For the next experiment, cost-function definitions beyond the pure least-square objective are applied to transition to more representative planning scenarios. This violates the formal equivalency expressed by \cref{eq:exp-value-f} such that scenario-free optimization is no longer equivalent to the scenario-based stochastic approach. Further, the impact of choosing other common sampling techniques using gridded scenarios (i.\,e., predetermined shifts over random samples) on the scenario-free approach are illustrated.

Two plans were optimized with the scenario-free approach exploiting a random and worst-case sampling technique. The worst-case scenarios were selected as described in \cref{subsection:MM_calc_prob_quantities}, including a total of \num{29} error scenarios. For compatibility, the sample size for the random sampling was also set to \num{29} scenarios. The probabilistic quantities were then computed on both sets. For additional comparison, a third plan was optimized with the stochastic approach using the same set of randomly sampled \num{29} error scenarios.

Subsequently, to asses the impact of the sample size, three additional sets of probabilistic quantities were computed with a sample size of \num{5}, \num{50} and \num{100} randomly sampled scenarios. For this sample size analysis, plans were optimized for the scenario-free algorithm only.

For all the optimized plans, robustness analysis was subsequently performed on a pool of \num{100} randomly sampled error scenarios.

\begin{figure}[H]
    \centering
    \resizebox{\textwidth}{!}{
        \includegraphics{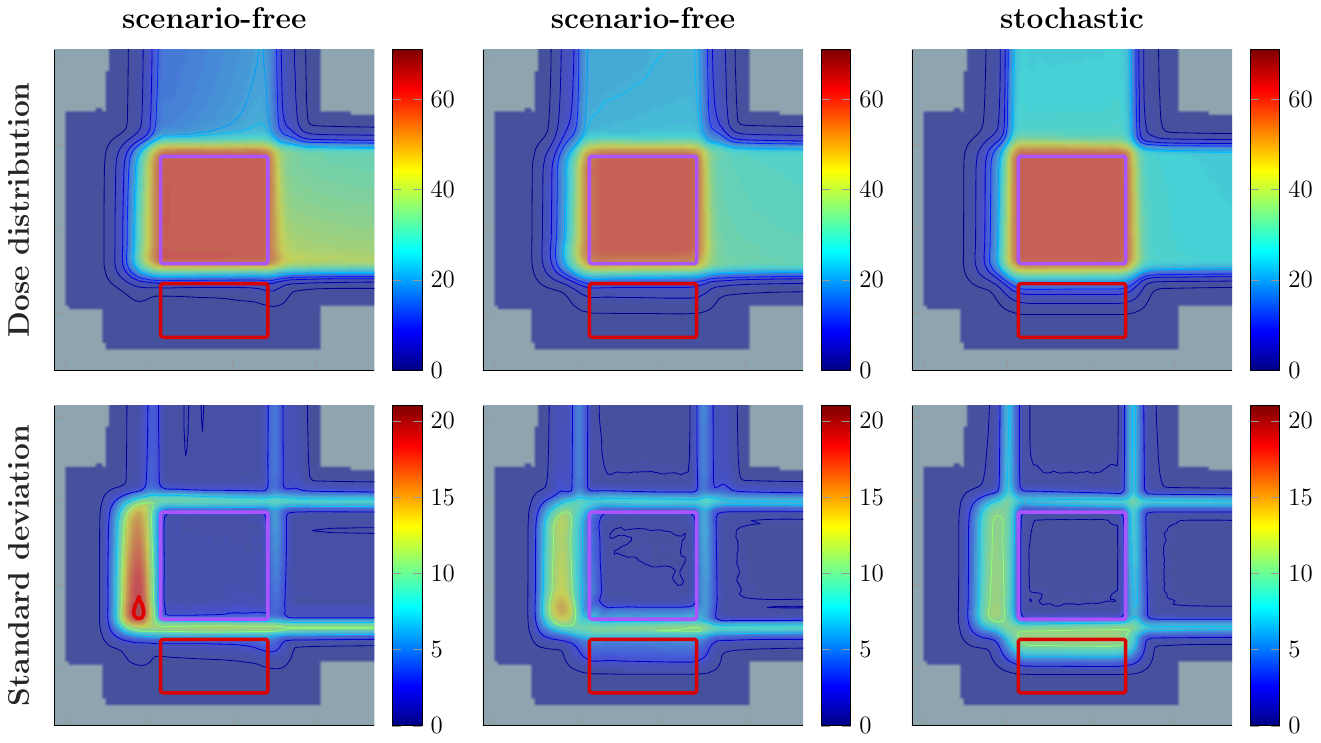}
    }
    \caption{Expected dose and SD distributions obtained with the different approaches and scenario sampling techniques described in \cref{subsection:MM_calc_prob_quantities}. 
    Top row: expected dose distributions obtained with the scenario-free algorithm and a randomly sampled set of scenarios (left), the scenario-free algorithm and a worst-case sampling approach (middle), the stochastic algorithm applied to the same set of randomly sampled scenarios (right). Bottom row: the corresponding standard deviation distributions. All colorbar values are reported in \si{\gray}.}
    \label{fig:distributions_sf_sf_stoch_rnd_wc}
\end{figure}

\Cref{fig:distributions_sf_sf_stoch_rnd_wc} reports the expected dose and SD distributions obtained with the different sampling techniques and optimization algorithms. As expected, the scenario-free and stochastic results are no longer matching, since the scenario-free approach uses direct variance minimization independent of the (expected) dose objectives used on the VOIs. Both the plans obtained with the scenario-free algorithm showed a lower SD distribution in the OAR compared to the stochastic optimization. This latter exhibits more balanced contribution of the two beams, as highlighted by the dose values in the entrance channels and the symmetric distribution of uncertainty.

The random-sampling plan applied to the scenario-free approach showed a better sparing of the OAR from uncertainty when compared to the worst-case sampling approach, and resulted in a higher distribution of dose and SD delivered by the \si{90}{\degree} proton field.

\begin{figure}[H]
    \begin{subfigure}[b]{0.5\textwidth}
        \resizebox{1.02\textwidth}{!}{
            \includegraphics{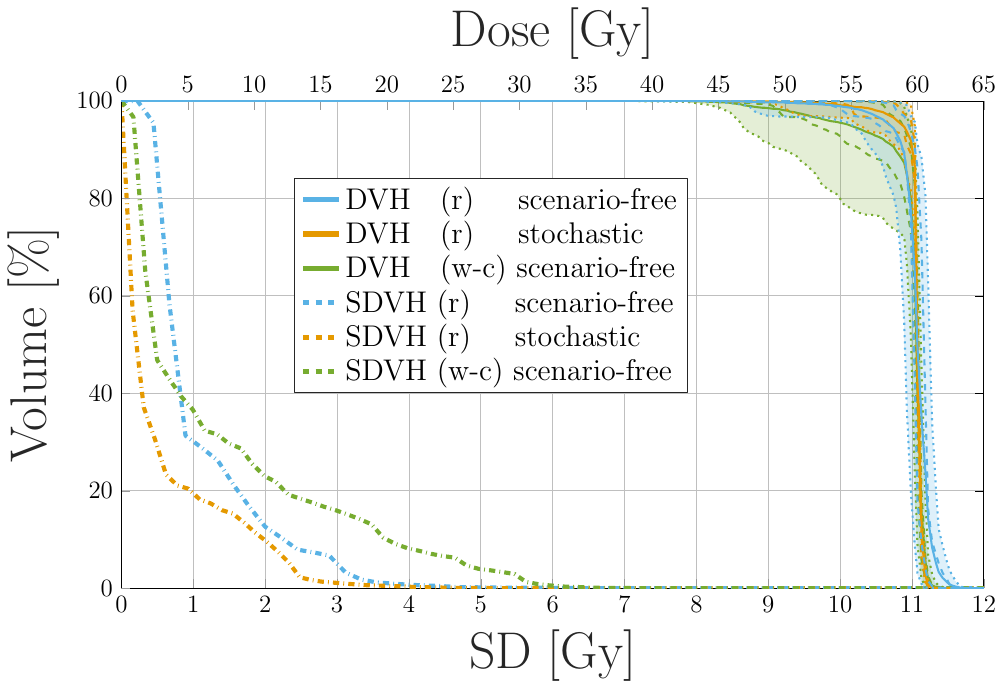}
        }
        \caption{Target DVHs and SDVHs}
        \label{fig:wc_5_to_100_panel_a}
    \end{subfigure}
    \begin{subfigure}[b]{0.5\textwidth}
        \resizebox{\textwidth}{!}{
            \includegraphics{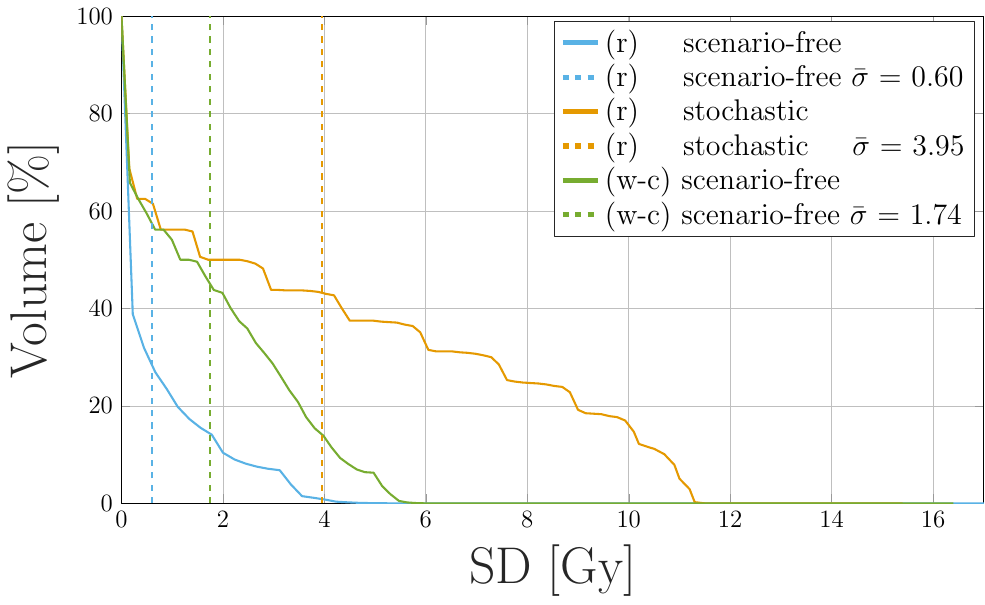}
        }
        \caption{OAR SDVHs}
        \label{fig:wc_5_to_100_panel_b}
    \end{subfigure}

    \begin{subfigure}[b]{0.5\textwidth}
        \resizebox{\textwidth}{!}{
            \includegraphics{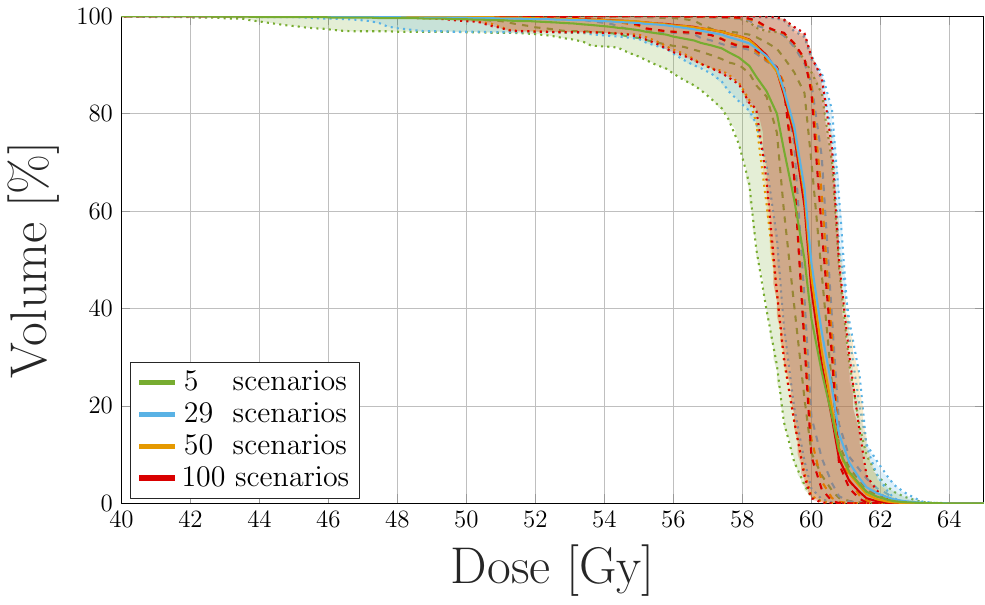}
        }
        \caption{Target DVHs}
        \label{fig:wc_5_to_100_panel_c}
    \end{subfigure}
    \begin{subfigure}[b]{0.5\textwidth}
        \resizebox{\textwidth}{!}{
            \includegraphics{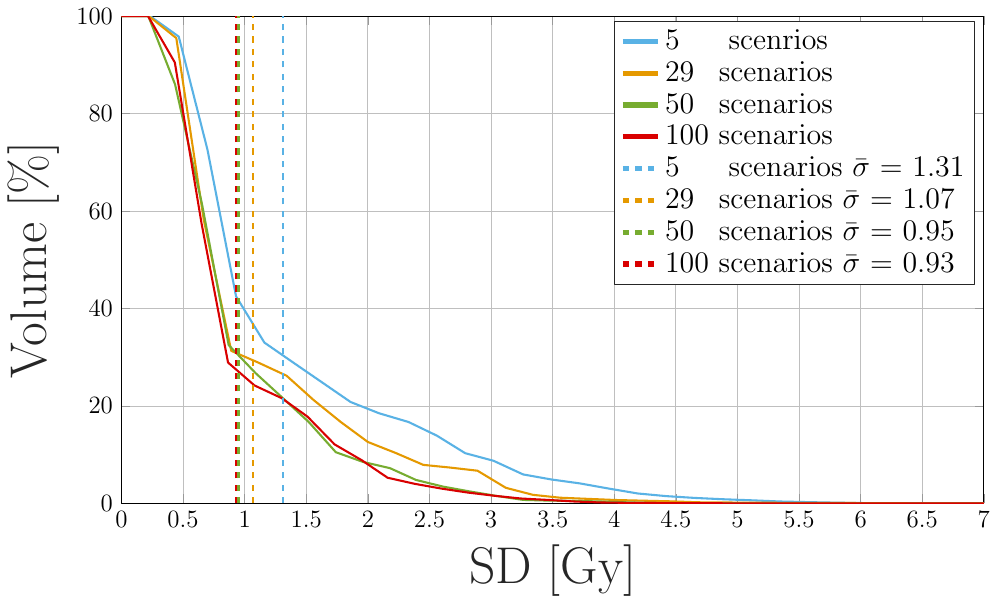}
        }
        \caption{Target SDVHs}
        \label{fig:wc_5_to_100_panel_d}
    \end{subfigure}
    \caption{Robustness analysis for the different sampling and optimization approaches. (a) DVHs and SDVHs for the target structure and (b) SDVHs for the OAR reported for both the sampling procedures applied to the scenario-free approach and the additional stochastic approach. Target (c) DVHs and (d) SDVHs for the scenario-free optimization when different sample sizes are used. For the reported DVHs, the solid line represents the DVH computed for the expected dose distribution while the dashed and dotted lines correspond to the \num{25}-\num{75} and the \num{5}-\num{95} percentiles of single-scenario DVHs distributions. The dotted vertical lines in the SDVH plot represent the mean standard deviation, i.e., the average value of the SD distribution.}
    \label{fig:wc_5_to_100_panel}
\end{figure}

\Cref{fig:wc_5_to_100_panel_a,fig:wc_5_to_100_panel_b} display DVHs and SDVHs for the optimized  plans.
The average SD values in the target structure resulted in close values of \SI{1.1}{\gray} and \SI{1.3}{\gray} for the random and worst-case sampling applied to the scenario-free approach respectively. The average SD value for the stochastic plan was  \SI{0.6}{\gray}. The shape of the SDVH, however, shows how the worst-case sampling allows for a limited amount of voxels to reach high SD values. The SDVH for the OAR is shifted to larger values for the scenario-free worst-case sampling, and the stochastic algorithm in order, as already highlighted by the distributions in \cref{fig:distributions_sf_sf_stoch_rnd_wc}.

The robustness analysis in \cref{fig:wc_5_to_100_panel_c,fig:wc_5_to_100_panel_d} collects the DVHs and SDVHs obtained with the scenario-free algorithm for different sizes of scenarios sample. While no significant difference in the distribution of scenario DVHs can be observed, the SDVHs present a clear trend of mean standard deviation reduction with the increase of sample size.

    \subsection{4D Box}
    As the scenario-free approach is expected to show its biggest benefit in high-dimensional uncertainty models, application to 4D planning was evaluated. Artificial target motion was initially simulated via multiple CT-phases of the previous box geometry. The complete set of scenarios includes thus both error scenarios and CT-phases.
Unless explicitly specified, for all 4D setups the target structure for robust optimization coincides with the CTV of each CT-phase. The target for the nominal optimization is instead always a PTV obtained by margin expansion of the ITV as indicated in \cref{tab:geom-phantoms}.
During robustness analysis instead, each distribution, DVH and SDVH is evaluated on the scenario-specific structures, i.e. the CTV is used as a target structure.

\begin{figure}[bth]
    \centering
    \begin{subfigure}[b]{0.49\textwidth}
        \resizebox{0.49\textwidth}{!}{
            \includegraphics{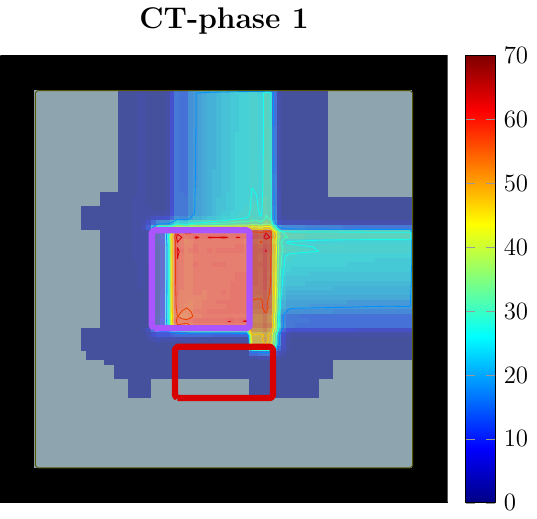}
        }
        \resizebox{0.49\textwidth}{!}{
            \includegraphics{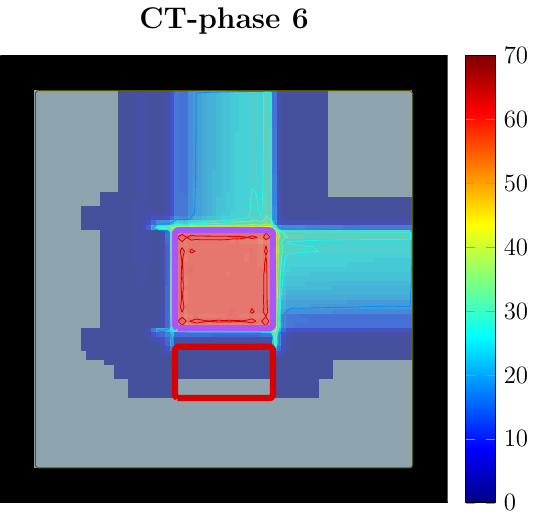}
        }

        \resizebox{0.49\textwidth}{!}{
            \includegraphics{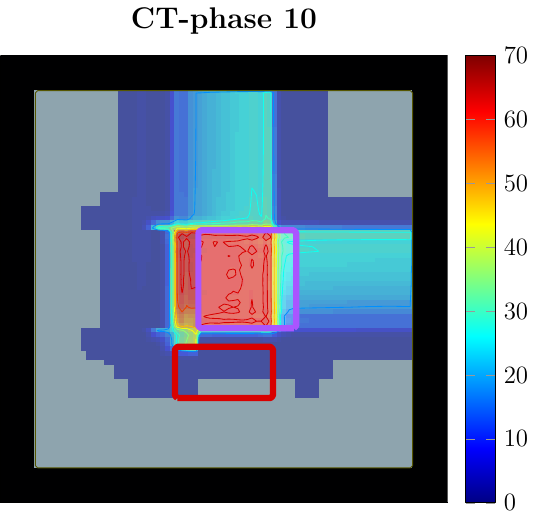}
        }
        \resizebox{0.49\textwidth}{!}{
            \includegraphics{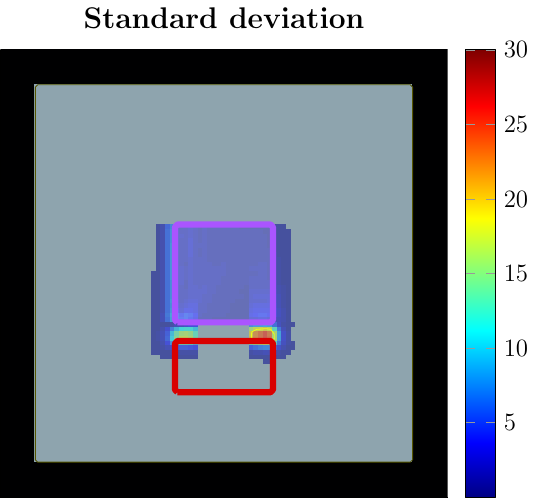}
        }
        \caption{Proton dose and SD distributions.}
        \label{fig:4D-Box-doseDist_a}
    \end{subfigure}
    \begin{subfigure}[b]{0.49\textwidth}
        \resizebox{0.49\textwidth}{!}{
            \includegraphics{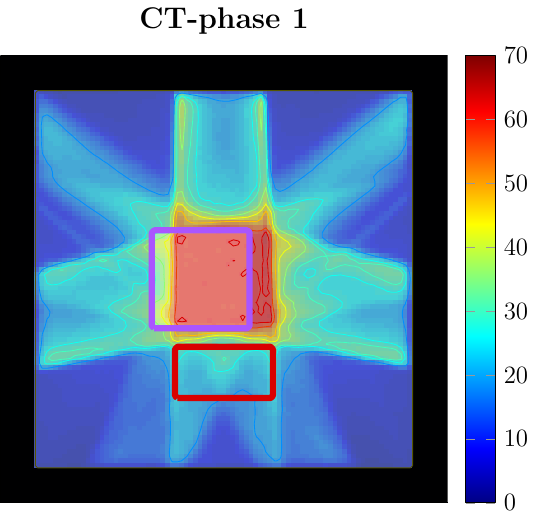}
        }
        \resizebox{0.49\textwidth}{!}{
            \includegraphics{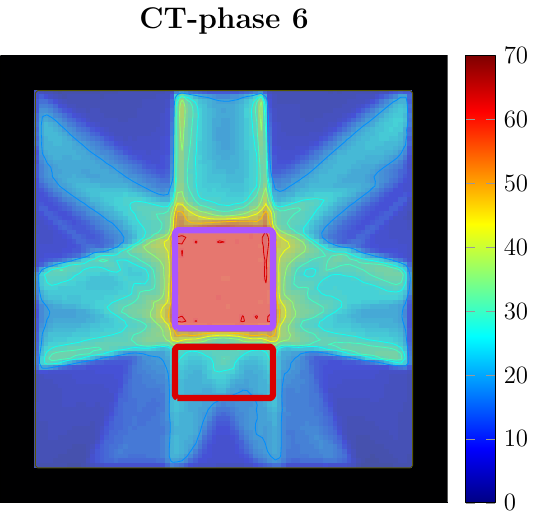}
        }

        \resizebox{0.49\textwidth}{!}{
            \includegraphics{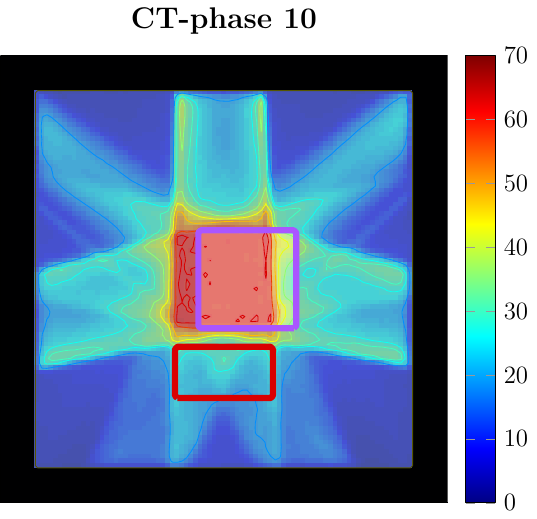}
        }
        \resizebox{0.49\textwidth}{!}{
            \includegraphics{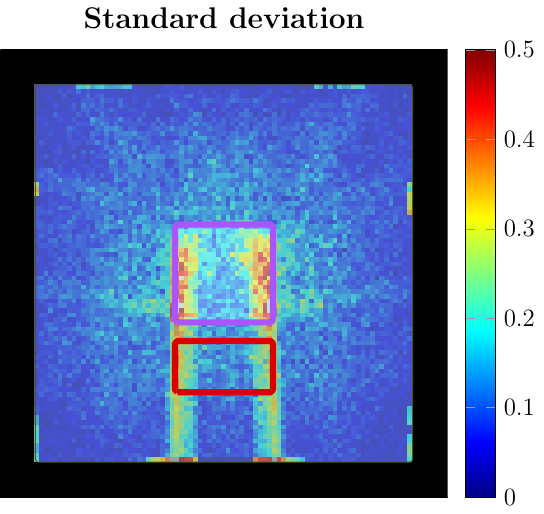}
        }
        \caption{Photons dose and SD distributions.}
        \label{fig:4D-Box-doseDist_v}
    \end{subfigure}
    \caption{Dose distributions obtained for the CT-phases at minimum (CT-phase \num{1}), nominal (CT-phase \num{6}) and maximum (CT-phase \num{10}) motion amplitude, together with the relative standard deviation distribution. The plans correspond to the proton (a) and photon (b) irradiation plans. The dose distributions are obtained as part of post-optimization robustness analysis performed on the nominally optimized distribution.}
    \label{fig:4D-Box-doseDist}
\end{figure}

In addition to the previous analysis, IMRT plans were also generated and allowed for the assessment of the algorithm performances for photon irradiation as well. \Cref{fig:4D-Box-doseDist} illustrates phase dose distributions and their standard deviation for a nominal optimized plan recalculated on the CT phases.

Three plans were optimized for each radiation modality applying the scenario-free, the traditional stochastic, and a nominal, margin-based algorithm. Least-square and squared-overdosing objectives were applied consistently to the target structure (CTV for the robust, ITV for the margin-based approaches) and OARs.

\begin{figure}[H]
    \centering
    \begin{adjustbox}{minipage=\linewidth,scale=1}
        \begin{subfigure}[b]{0.45\textwidth}
            \resizebox{\textwidth}{!}{
                \includegraphics{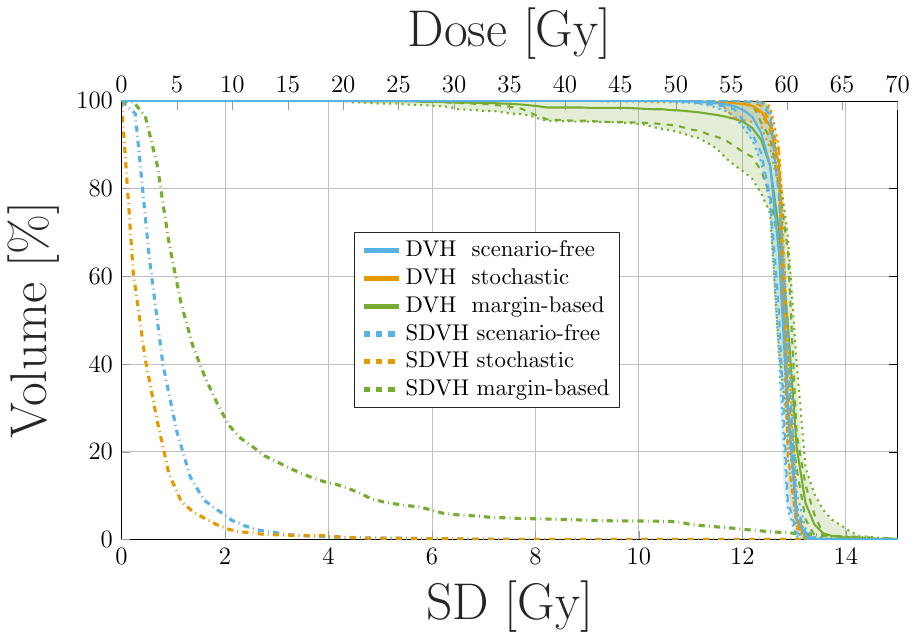}
            }
        \caption{Proton Target DVHs and SDVHs}
        \label{fig:4D-Box-DVH-SDVH_a}
        \end{subfigure}
        \hfill
        \begin{subfigure}[b]{0.45\textwidth}
                \resizebox{\textwidth}{!}{
                    \includegraphics{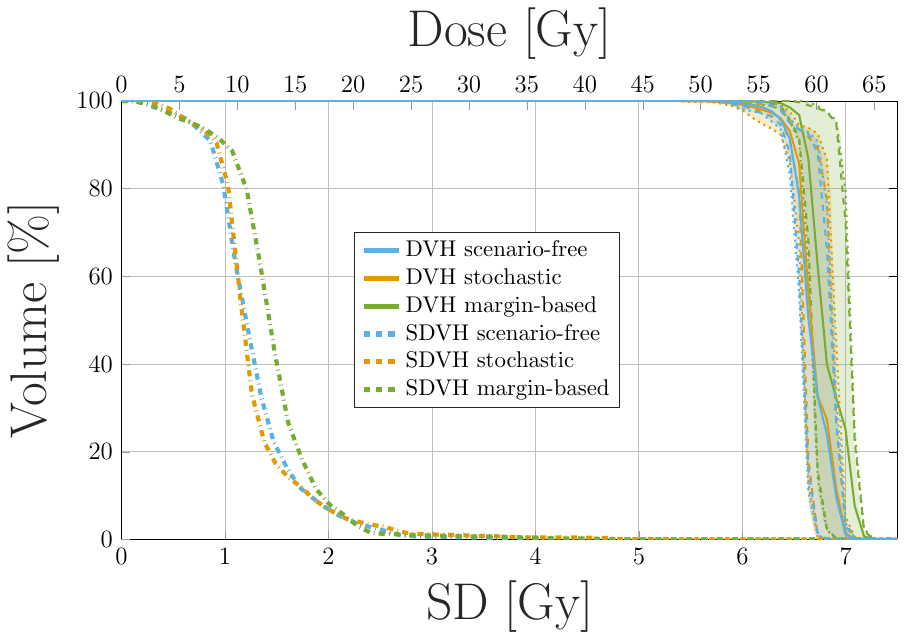}
                }
            \caption{Photons Target DVHs and SDVHs}
            \label{fig:4D-Box-DVH-SDVH_b}
        \end{subfigure}

        \begin{subfigure}[b]{0.45\textwidth}
                \resizebox{\textwidth}{!}{
                    \includegraphics{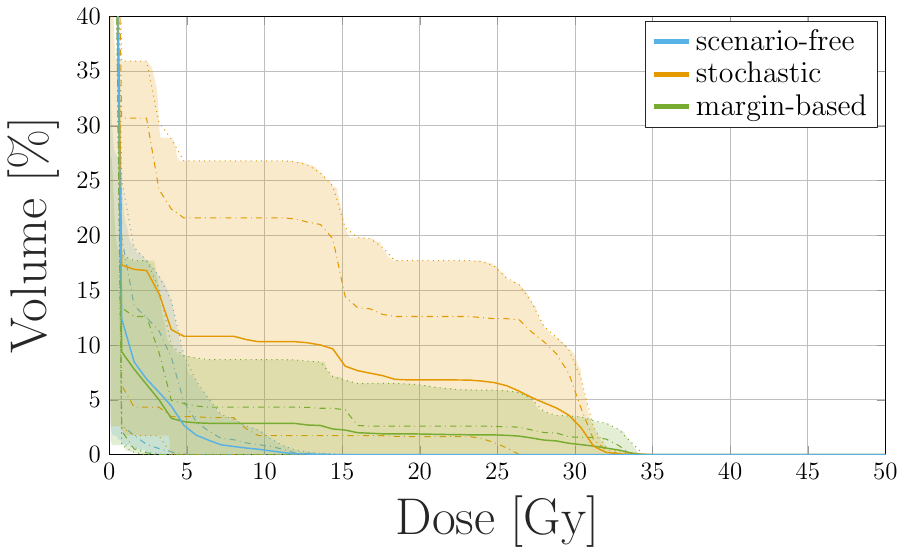}
                }
            \caption{Proton OAR DVHs}
            \label{fig:4D-Box-DVH-SDVH_c}
        \end{subfigure}
        \hfill
        \begin{subfigure}[b]{0.45\textwidth}
            \resizebox{\textwidth}{!}{
                \includegraphics{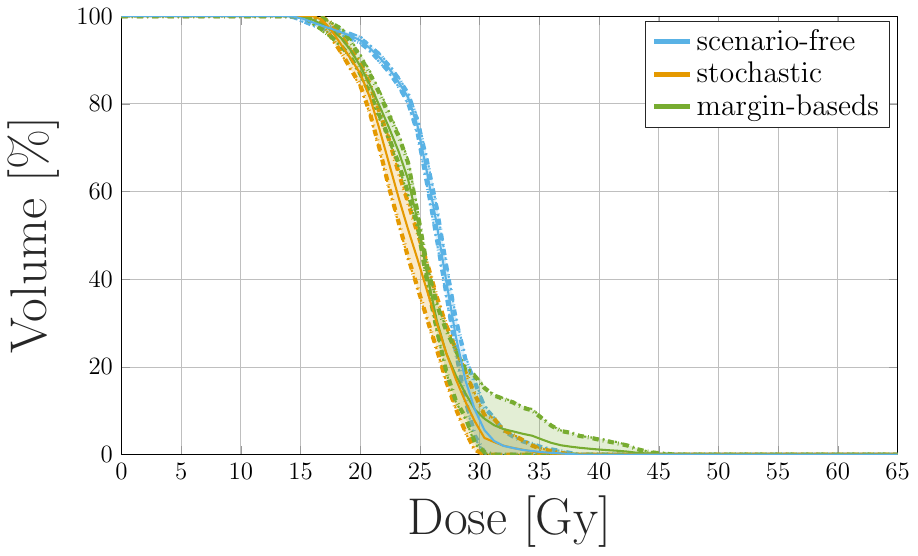}
            }
            \caption{Photons OAR DVHs}
            \label{fig:4D-Box-DVH-SDVH_d}
        \end{subfigure}

        \begin{subfigure}[b]{0.45\textwidth}
            \resizebox{\textwidth}{!}{
                \includegraphics{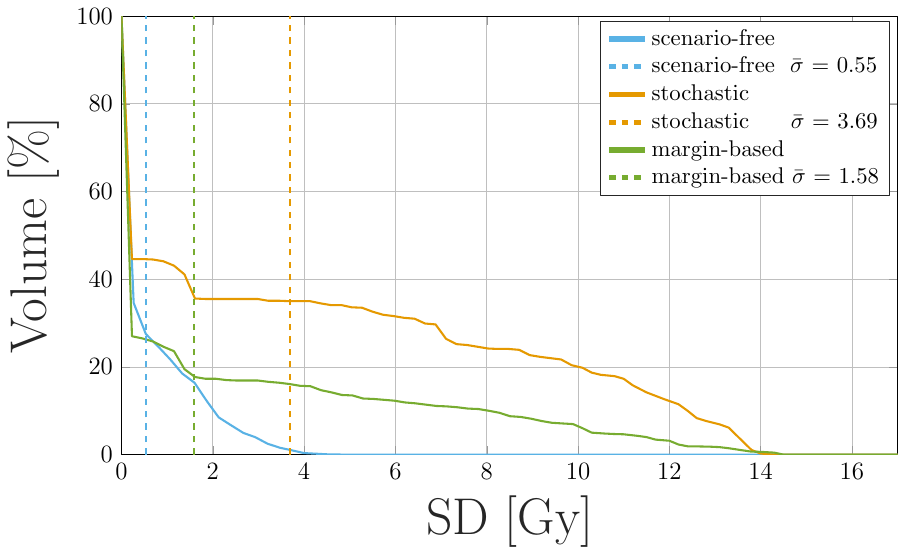}
            }
            \caption{Protons OAR SDVHs}
            \label{fig:4D-Box-DVH-SDVH_c}
        \end{subfigure}
        \hfill
        \begin{subfigure}[b]{0.45\textwidth}
            \resizebox{\textwidth}{!}{
                \includegraphics{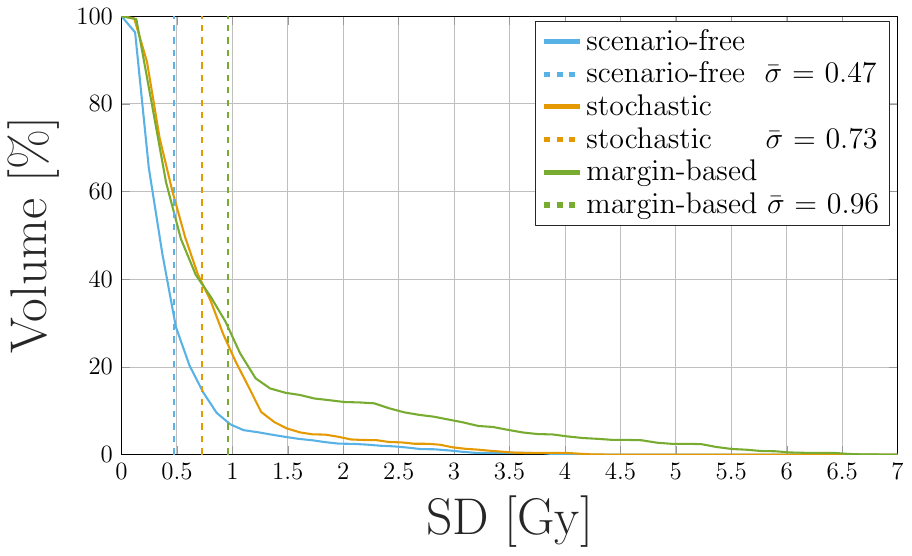}
            }
            \caption{Photons OAR SDVHs}
            \label{fig:4D-Box-DVH-SDVH_f}
        \end{subfigure}
    \end{adjustbox}
    \caption{Dosimetric and robustness analysis performed for the photon and proton plans optimized with the scenarios-free, traditional stochastic and margin-based methods for the 4D box-shaped phantom.
    Top: DVH and SDVH for the target structure for the proton (a) and photons (b) irradiation plans. Middle: DVHs for the OAR for proton (c) and photons (d) irradiaiton. Bottom: SDVHs for the OAR for proton (e) and photons (f) irradiaiton. For the reported DVHs, the solid line represents the DVH computed for the expected dose distribution while the dashed and dotted lines correspond to the \num{25}-\num{75} and the \num{5}-\num{95} percentiles of single-scenario DVHs distributions. The dotted vertical lines in the SDVH plot represent the mean standard deviation, i.e., the average value of the SD distribution.}
   \label{fig:4D-Box-DVH-SDVH}
\end{figure}

Robustness analysis was performed on the optimized plans and comparison between DVHs and SDVHs is reported in \cref{fig:4D-Box-DVH-SDVH}. As a general remark, the obtained proton plan tends to achieve comparable target coverage to the photon plan. The observed dose delivered to the OAR was consistently lower for all three optimization methods for the proton plan, as opposed to the photon irradiation, while the opposite is true for the SDVH curves. Even though comparison between radiation modalities goes beyond the scope of this analysis, the expected behavior is observed, with the proton dose distribution being more conformal to the target and, concurrently, more sensitive to uncertainty than the photon distribution.

For both modalities the robustness in target coverage was higher for both the robust algorithms when compared to the margin based optimization. This feature is highlighted by the SDVH curves depicted in \cref{fig:4D-Box-DVH-SDVH_a,fig:4D-Box-DVH-SDVH_d}. A more robust target coverage could be achieved with the nominal optimization by increasing the ITV margin, at the cost of increasing the integral dose and with the risk of delivering higher doses to the surrounding OARs as well.

For the proton plan, the scenario-free approach was capable of achieving lower dose and SD within the OAR when compared to the traditional stochastic algorithm, while maintaining comparable target coverage and robustness. This may be attributed to the explicit variance reduction objectives introduced by the scenario-free algorithm; since the applied cost-functions are not limited to pure least-square, higher penalization of the variance is expected.

\begin{figure}[H]
    \centering
    \begin{subfigure}[b]{0.45\textwidth}
        \resizebox{\textwidth}{!}{
            \includegraphics{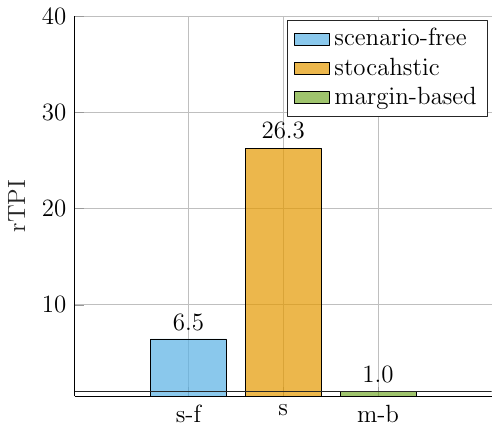}
        }
        \caption{Proton plan rTPI}
        \label{fig:4D-box-times_a}
    \end{subfigure}
    \begin{subfigure}[b]{0.45\textwidth}
        \resizebox{\textwidth}{!}{
            \includegraphics{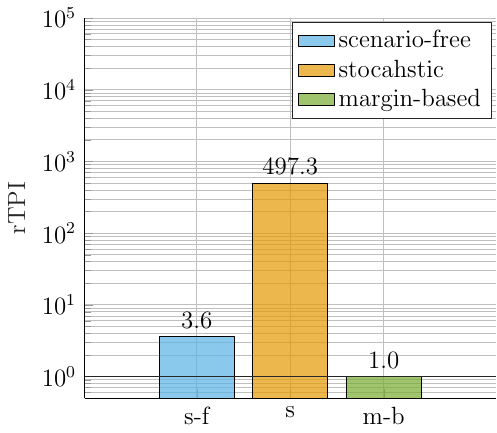}
        }
        \caption{Photon plan rTPI}
        \label{fig:4D-box-times_b}
    \end{subfigure}
    \caption{Relative time per iteration (rTPI) for the proton (left) and photons (right) 4D robust plans. The abscissa labels refer respectively to the scenario-free (s-f), stochastic (s) and margin-based (m-b) approaches. The photon rTPI is reported in logarithmic scale for clarity.}
    \label{fig:4D-box-times}
\end{figure}

The runtime efficiency for the three algorithms was assessed in terms of relative TPI (rTPI). Considering the TPI for the margin-based plan as a reference value, rTPI for the scenario-free and traditional stochastic optimization algorithms are reported in \cref{fig:4D-box-times} for both radiation modalities.

For both radiation modalities, the observed rTPI for the scenario-free algorithm was significantly lower than the stochastic approach. Despite the high number of bixels exploited by the proton plan (in the order of \num{23e3} bixels, against \num{2.7e3} for the photon plan), the number of non-zero elements stored in the sparse single-scenario dose influence matrices was limited by the finite range of the dose distribution. The discrepancy between the scenario-free and the stochastic rTPI is thus enhanced by the memory usage for the photon irradiation.

\begin{figure}[H]
    \centering
    \begin{subfigure}[b]{0.49\textwidth}
        \hspace*{-0.5cm}
        \resizebox{\textwidth}{!}{
            \includegraphics{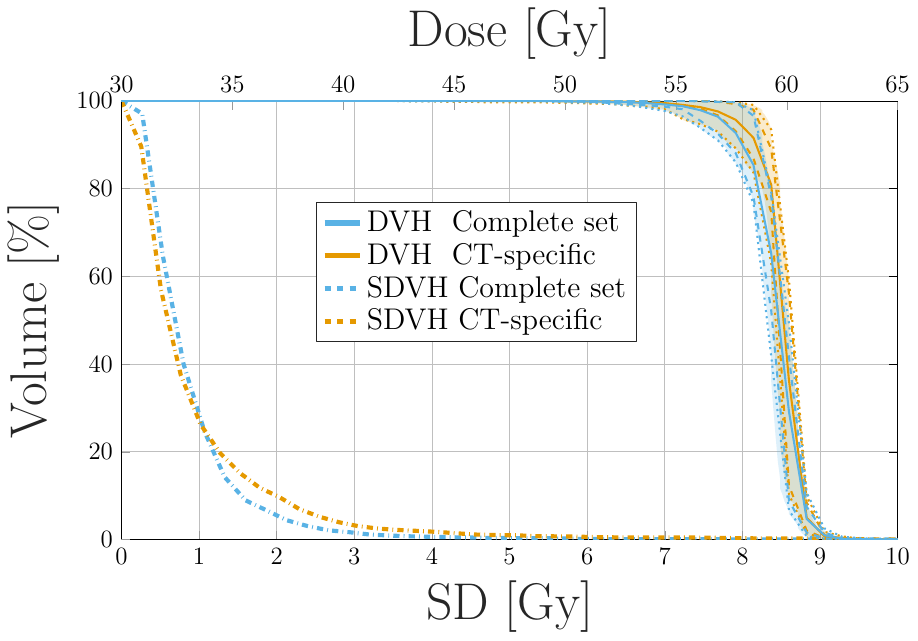}
        }
        \caption{Proton plan}
        \label{fig:4D-Box-phase-vs-all_a}
    \end{subfigure}
    \begin{subfigure}[b]{0.49\textwidth}
        \resizebox{\textwidth}{!}{
            \includegraphics{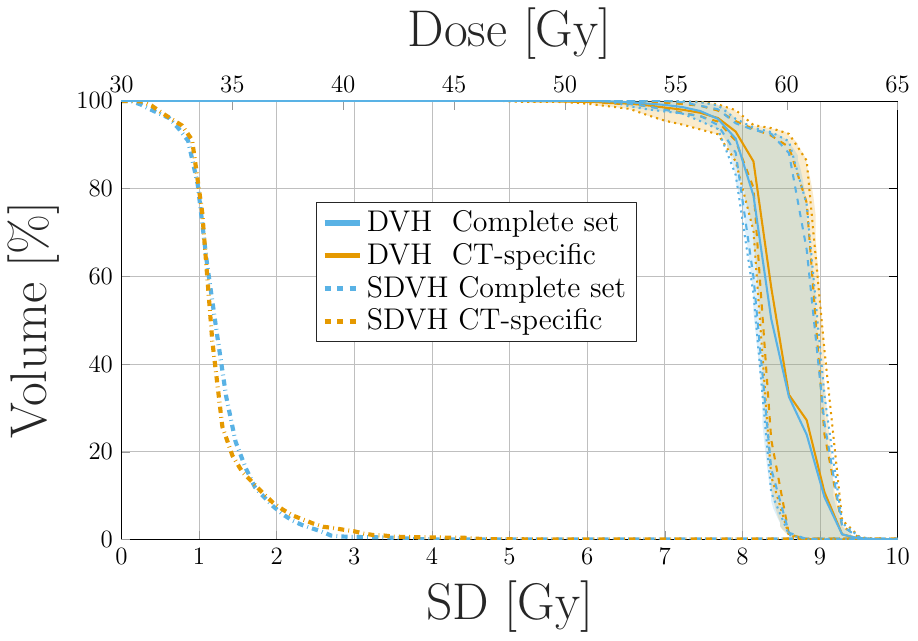}
        }
        \caption{Photon plan}
        \label{fig:4D-Box-phase-vs-all_b}
    \end{subfigure}
    \caption{Dosimetric and robustness analysis performed for the two calculation strategies described in \cref{subsection:MM_calc_prob_quantities} for the 4D scenario accumulation.
    DVHs and SDVHs are obtained for the target structure with the two calculation approaches for proton (a) and photon (b) irradiation.For the reported DVHs, the solid line represents the DVH computed for the expected dose distribution while the dashed and dotted lines correspond to the \num{25}-\num{75} and the \num{5}-\num{95} percentiles of single-scenario DVHs distributions.}
   \label{fig:4D-Box-phase-vs-all}

\end{figure}

\Cref{fig:4D-Box-phase-vs-all} reports a comparison between plans optimized with the scenario-free approach. The two plans correspond to the different modes described in section \cref{subsection:MM_calc_prob_quantities} to compute and combine the probabilistic quantities on multiple CT-phases. Both plans achieve comparable target coverage and robustness.

    \subsection{4D Patient Case}
    The characterization of the scenario-free algorithm performances was also extended to more clinically realistic patient plans. For this analysis a 4D lung cancer patient dataset comprising of \num{30} scenarios in total was selected. This included \num{10} CT-phases modeling the breathing motion and \num{3} error scenarios for each phase as described in \cref{subsection:Treatment_plans}. Even in this case both protons and photons were applied as radiation modalities.

\begin{figure}[hbt]
    \resizebox{\textwidth}{!}{
        \includegraphics{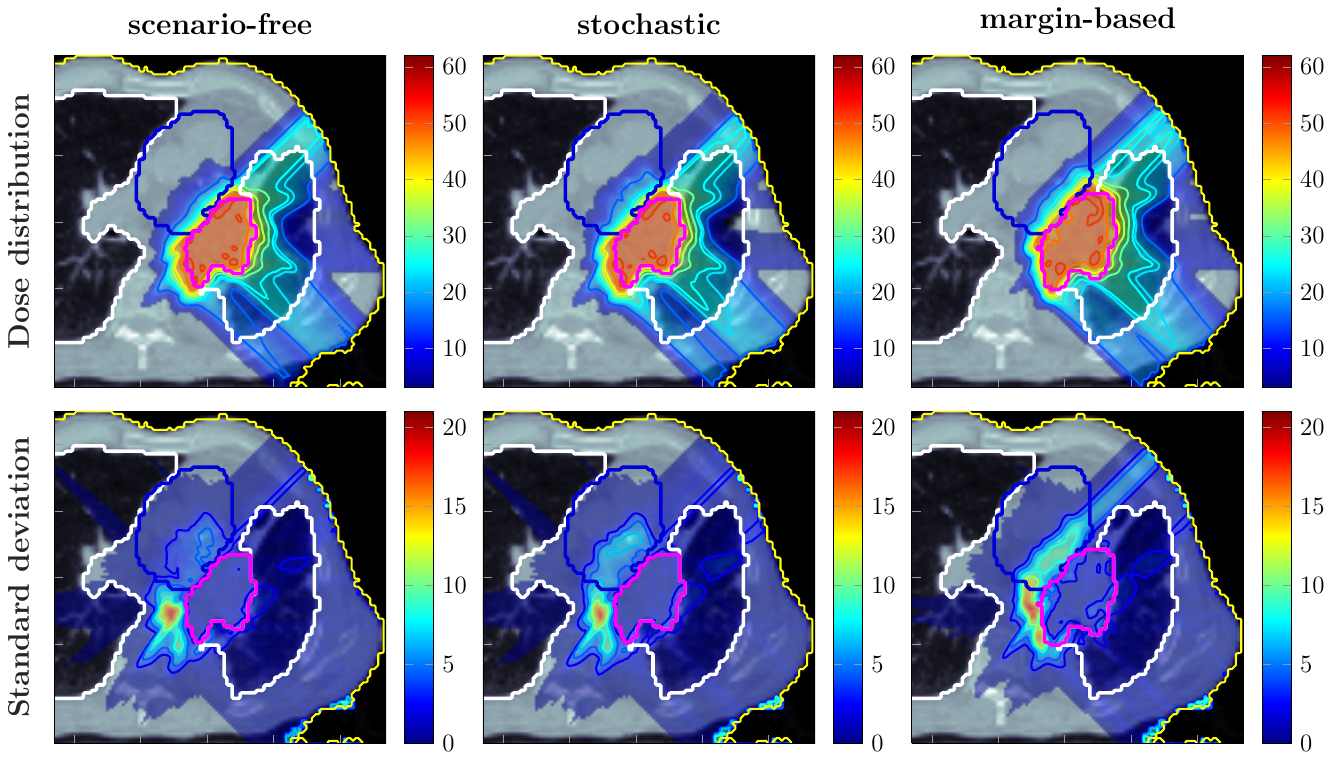}
    }
    \caption{Comparison between proton dose distributions (top row) and SD distributions (bottom row), both reported in \si{\gray}, obtained with the scenario-free (left), the stochastic (middle) and margin-based (right) optimization algorithms. The contoured structures correspond to the lungs (white), the heart (blue), and the target structure (purple). The target structure coincides with the CTV for the robust optimization algorithms, and with the PTV for the margin-based plan.}
    \label{fig:P1-distributions-protons}
\end{figure}

\begin{figure}[hbtp]
    \resizebox{\textwidth}{!}{
        \includegraphics{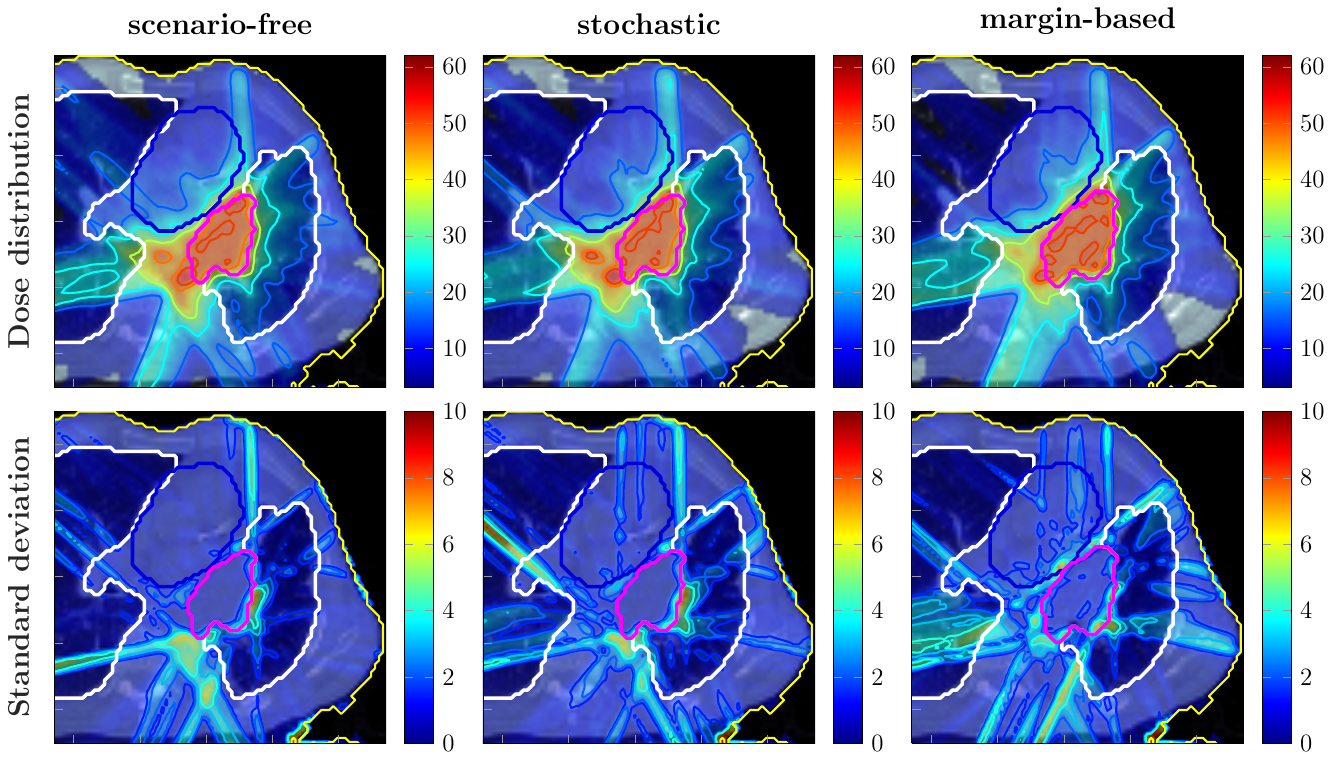}
    }
    \caption{Comparison between photon dose distributions (top row) and standard deviation distributions (bottom row), both reported in \si{\gray}, obtained with the scenario-free (left), the stochastic robust (middle) and margin-based (right) optimization algorithms. The contoured structures correspond to the lungs (white), the heart (blue), and the target structure (purple). The target structure coincides with the CTV for the robust optimization algorithms, and with the PTV for the margin-based plan. }
    \label{fig:P1-distributions-photons}
\end{figure}

\Cref{fig:P1-distributions-protons,fig:P1-distributions-photons} collect examples of dose and standard deviation distributions obtained for the \num{3}-fields proton and the \num{7}-fields photons plans respectively, applying all three optimization approaches. The margin-based plan shows the highest SD within the target and the major OARs. On the contrary, the lowest values were observed for the scenario-free approach. 

\begin{figure}[]
    \centering
    \begin{adjustbox}{minipage=\linewidth,scale=1}
        \begin{subfigure}[b]{0.49\textwidth}
            \resizebox{\textwidth}{!}{
                \includegraphics{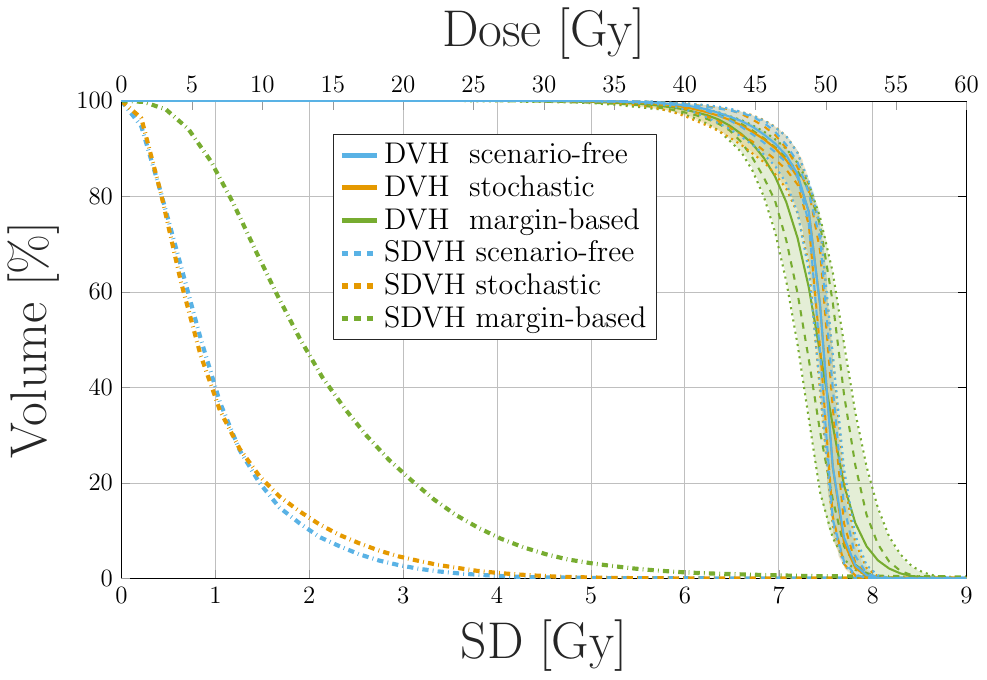}
            }
        \caption{Proton Target DVHs and SDVHs}
        \label{fig:P1_proton_DVHs_a}
        \end{subfigure}
        \hfill
        \begin{subfigure}[b]{0.49\textwidth}
            \resizebox{\textwidth}{!}{
                \includegraphics{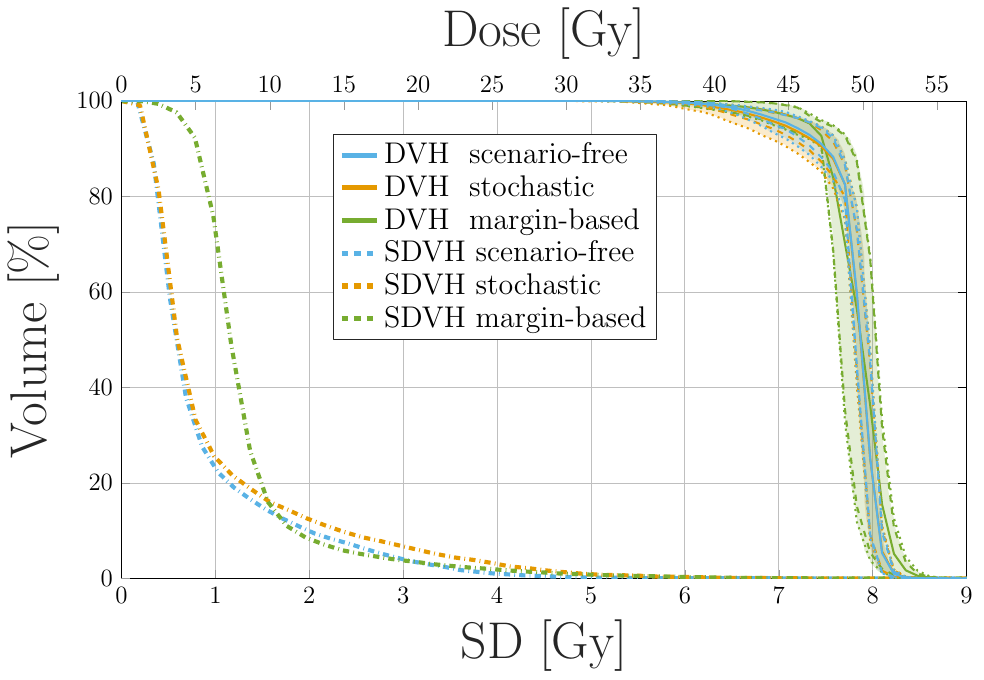}
            }
            \caption{Photons Target DVHs and SDVHs}
            \label{fig:P1_proton_DVHs_b}
        \end{subfigure}
    
        \begin{subfigure}[b]{0.49\textwidth}
                \resizebox{\textwidth}{!}{
                    \includegraphics{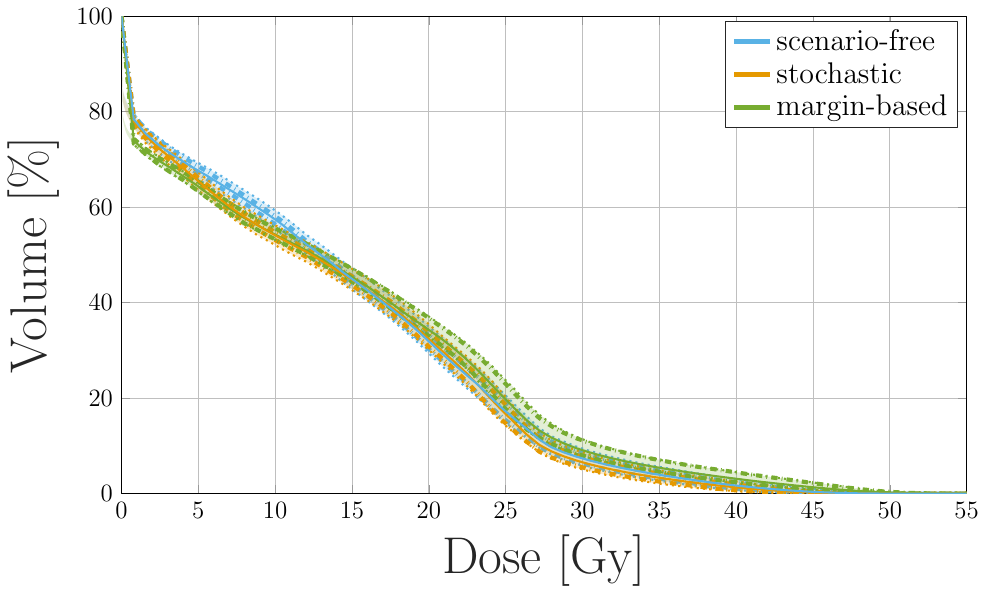}
                }
            \caption{Proton Lung DVHs}
            \label{fig:P1_proton_DVHs_c}
        \end{subfigure}
        \hfill
        \begin{subfigure}[b]{0.49\textwidth}
            \resizebox{\textwidth}{!}{
                \includegraphics{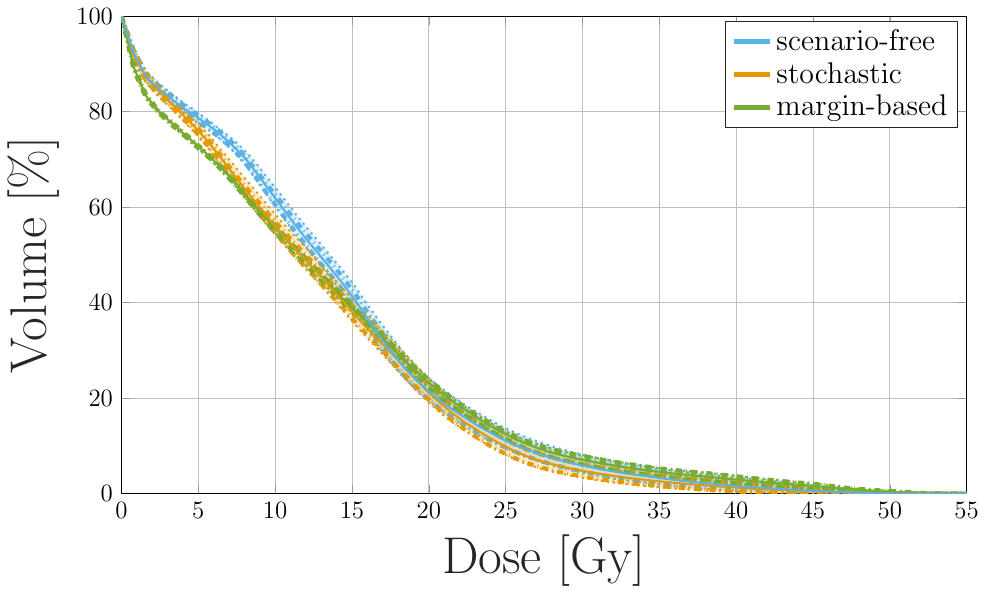}
            }
            \caption{Photons Lung DVHs}
            \label{fig:P1_proton_DVHs_d}
        \end{subfigure}

        \begin{subfigure}[b]{0.49\textwidth}
                \resizebox{\textwidth}{!}{
                    \includegraphics{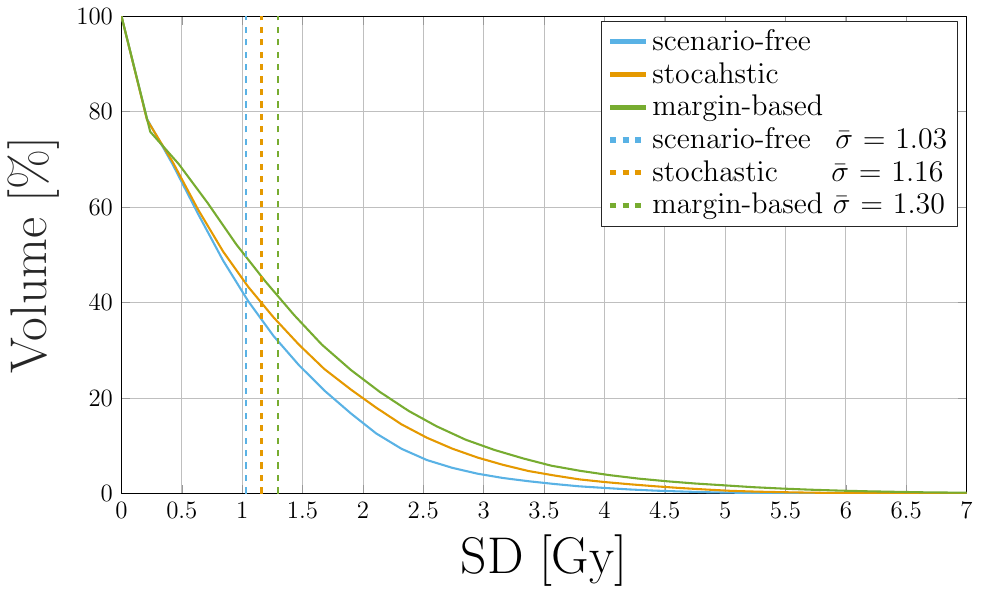}
                }
            \caption{Protons Lung SDVHs}
            \label{fig:P1_proton_DVHs_e}
        \end{subfigure}
        \hfill
        \begin{subfigure}[b]{0.49\textwidth}
            \resizebox{\textwidth}{!}{
                \includegraphics{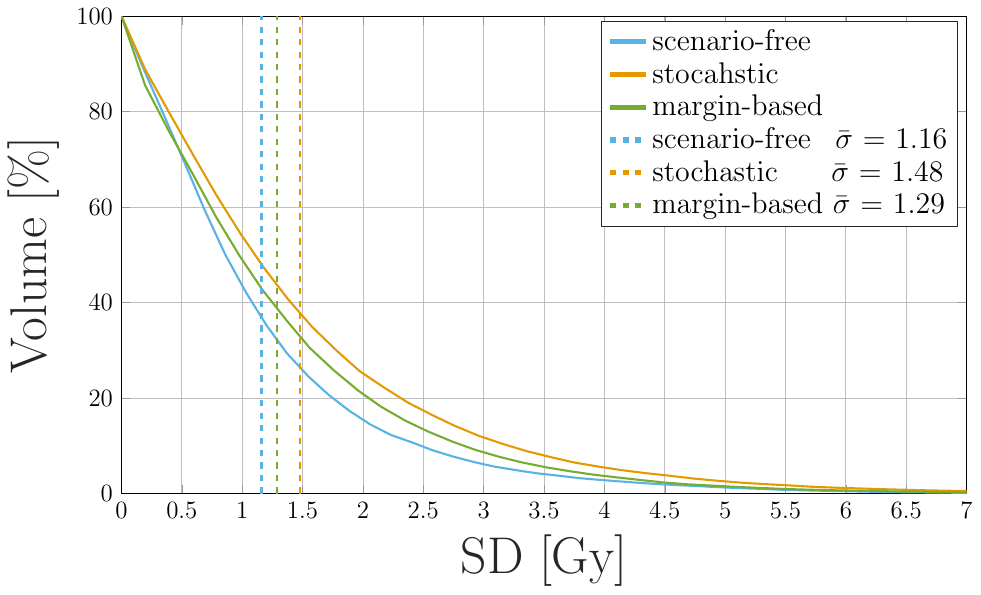}
            }
            \caption{Photons Lung SDVHs}
            \label{fig:P1_proton_DVHs_f}
        \end{subfigure}
    \end{adjustbox}
    \caption{Dosimetric and robustness analysis performed for the photon and proton plans optimized with the scenarios-free, traditional stochastic and margin-based methods.
    Top: DVH and SDVH for the target structure for the proton (a) and photons (b) irradiation plans. Middle: DVHs for the lung for proton (c) and photons (d) irradiaiton. Bottom: SDVHs for the lung for proton (e) and photons (f) irradiaiton. For the reported DVHs, the solid line represents the DVH computed for the expected dose distribution while the dashed and dotted lines correspond to the \num{25}-\num{75} and the \num{5}-\num{95} percentiles of single-scenario DVHs distributions. The dotted vertical lines in the SDVH plot represent the mean standard deviation, i.e., the average value of the SD distribution.}
   \label{fig:P1_proton_DVHs}
\end{figure}

Robustness analysis was performed including the dose distributions for all the error scenarios and optimization approaches. Results are collected and represented through DVHs and SDVHs in \cref{fig:P1_proton_DVHs}. In this case, the left lung was reported as a representative OAR.

The DVHs for the target structure reported in \cref{fig:P1_proton_DVHs_a} for the proton plan show an improved and more robust target coverage for the scenario-free approach. The distribution of single scenario DVHs clusters around the dose prescription with a narrow distribution for both robust algorithms. This is also reflected by the shift of the SDVH curve toward lower values of standard deviation.

\Cref{fig:P1_proton_DVHs_c,fig:P1_proton_DVHs_e} report the DVH and SDVH lines for the same plan and the lung. In this case, the scenario free-algorithm seemingly tends to slightly enhance the sparing of the organ from high doses and simultaneously reduce the standard deviation. The mean lung dose observed for the scenario-free, stochastic and margin-based plans was respectively \SI{13.3 \pm 0.5}{\gray}, \SI{16.6 \pm 0.5}{\gray} and \SI{13.2 \pm 0.28}{\gray}, 
resulting thus in a comparable value for the scenario-free and margin-based approaches, while slightly higher for the stochastic plan.

The same analysis was performed for the photon plan and is also reported in \cref{fig:P1_proton_DVHs_b,fig:P1_proton_DVHs_d,fig:P1_proton_DVHs_f}. Comparable target coverage was achieved by the robust optimization algorithms while the distribution of standard deviation within the target is reduced with respect to the margin-based approach. For this plan the DVH distributions for the lung showed less variability over the three methods.

\begin{figure}[H]
    \begin{subfigure}[b]{0.49\textwidth}
        \hspace*{0.3cm}
        \resizebox{\textwidth}{!}{
            \includegraphics{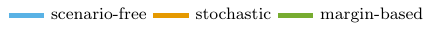}
        }

        \resizebox{0.49\textwidth}{!}{
            \includegraphics{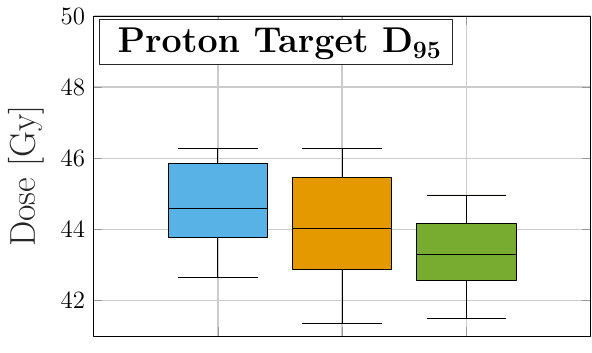}
        }
        \resizebox{0.49\textwidth}{!}{
            \includegraphics{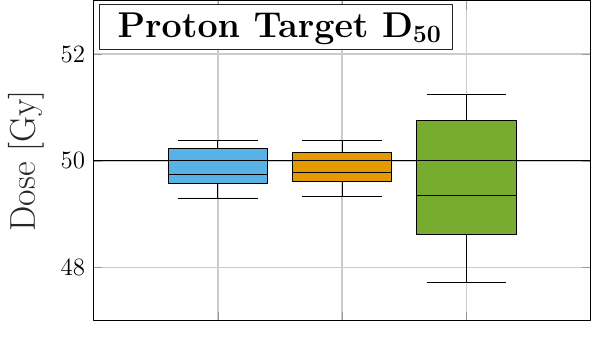}
        }

        \resizebox{0.49\textwidth}{!}{
            \includegraphics{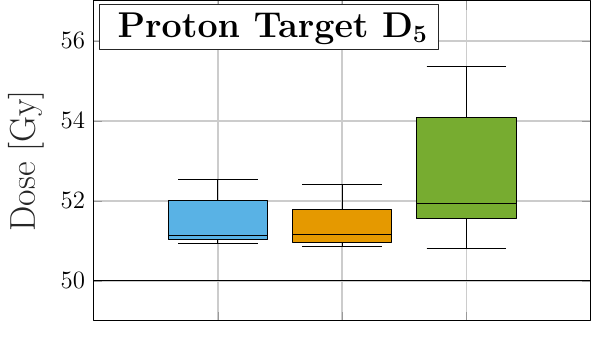}
        }
        \resizebox{0.49\textwidth}{!}{
            \includegraphics{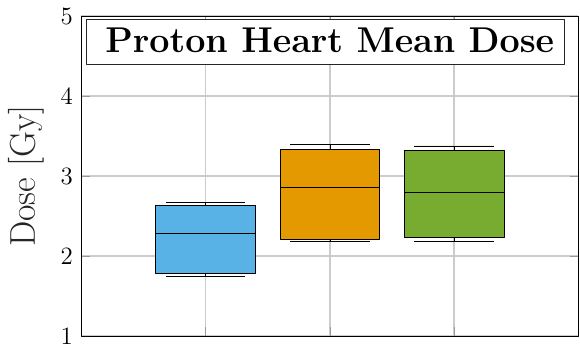}
        }
        \caption{Proton DVH points and heart mean dose}
    \end{subfigure}
    \hfill
    \begin{subfigure}[b]{0.49\textwidth}
        \hspace*{0.3cm}
        \raisebox{0.2cm}{
            \resizebox{\textwidth}{!}{
                \includegraphics{Legend.pdf}
            }
        }
        \resizebox{0.49\textwidth}{!}{
            \includegraphics{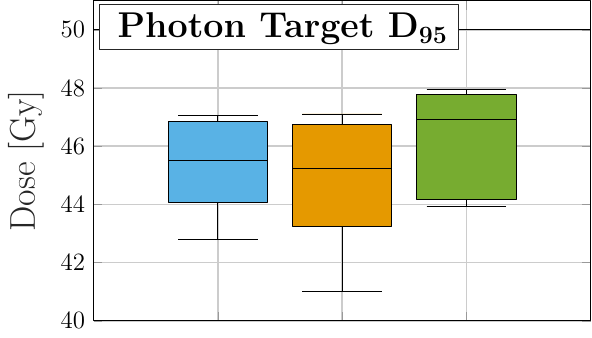}
        }
        \resizebox{0.49\textwidth}{!}{
            \includegraphics{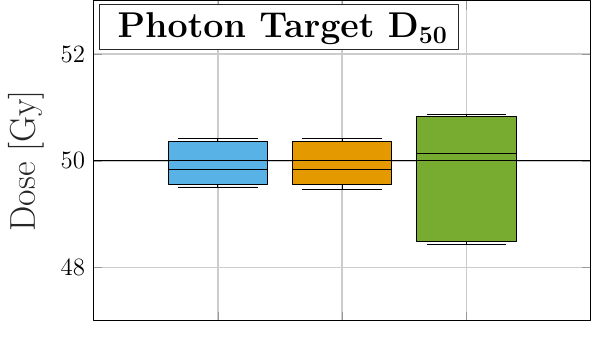}
        }

        \resizebox{0.49\textwidth}{!}{
            \includegraphics{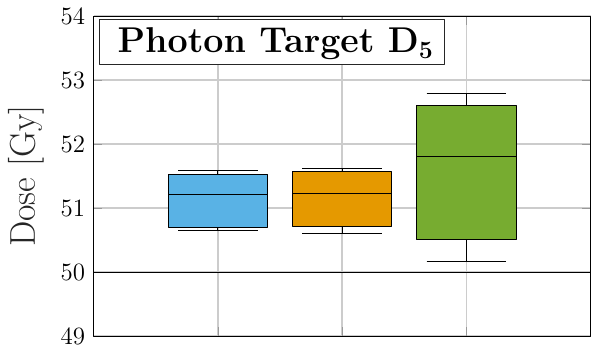}
        }
        \resizebox{0.49\textwidth}{!}{
            \includegraphics{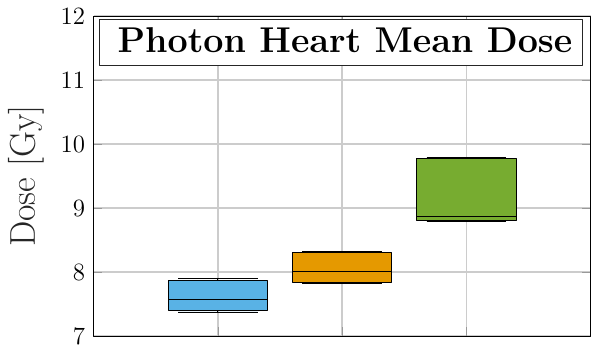}
        }
        \caption{Photon DVH points and heart mean dose}
    \end{subfigure}
    \caption{DVH points collected for the proton (a) and photon plan (b). The metrics' values are distributed over the set of error scenarios and reported for all three applied optimization methods. The solid line at \SI{50}{\gray} represents the dose prescription. Reported are also the heart mean dose for both modalities.}
    \label{fig:P1_BoxPlots}

\end{figure}

\Cref{fig:P1_BoxPlots} collects selected DVH points for the target structure and the mean dose values for the heart. The scenario-free and the traditional stochastic algorithm can achieve better target conformity. Compared to the nominal margin-based optimization, the box plots for the $D_{95}$, $D_{50}$ and $D_{5}$ values for these two algorithms cluster closer to the dose prescription value of \SI{50}{\gray}. The quartile boundaries are also closer to the median value, reflecting the behavior displayed by the SDVH curves in \cref{fig:P1_proton_DVHs}. The observed mean heart doses are also lower for the scenario-free approach in both radiation modalities.

\begin{figure}[hbtp]
    \begin{subfigure}{0.5\textwidth}
        \resizebox{\textwidth}{!}{
            \includegraphics{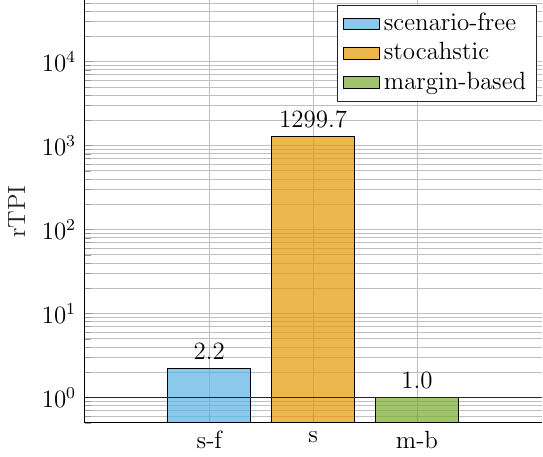}
        }
        \caption{Proton plan rTPI}
    \end{subfigure}
    \begin{subfigure}{0.5\textwidth}
        \resizebox{\textwidth}{!}{
            \includegraphics{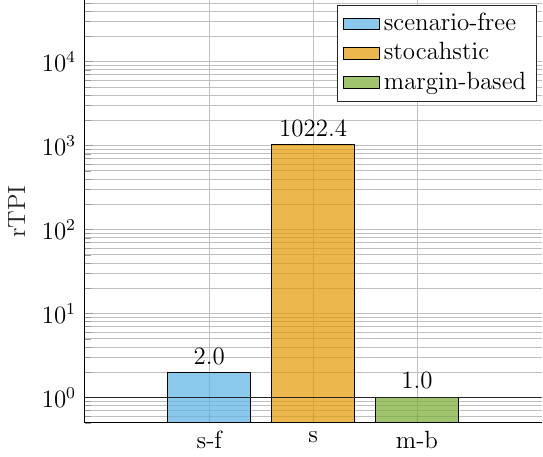}
        }
        \caption{Photon plan rTPI}
    \end{subfigure}
    \caption{rTPI for the proton (left) and photon (right) lung cancer patient plan. The abscissa labels refer respectively to the scenario-free (s-f), stochastic (s) and margin-based (m-b) approaches.}
    \label{fig:P1_times}
\end{figure}

The relative time per iteration results are reported in \cref{fig:P1_times} and show the same trend as observed for the 4D-box phantom in \cref{fig:4D-box-times}. However, the number of non-zero elements stored for the single-scenario dose influence matrices is enhanced here by the higher dimension of the patient CT, making the repeated evaluation of individual scenario distributions even more computationally demanding. The resulting discrepancy in rTPI between the different optimization strategies is thus more pronounced in a clinically realistic setup.

    \renewcommand{\subfolder}{./Figures/Patient_2/tikz}
For the second 4D patient, the number of scenarios was increased to \num{100}, combining \num{10} range and shift error scenarios for each of the \num{10} CT phases.
Multiple cost-functions were defined as reported in (\cref{tab:cost-fun}). Only the scenario-free and the margin-based approach were applied. Memory limitations in the available hardware prevented the application of the traditional stochastic optimization algorithm with such an high number of scenarios.

\begin{figure}[H]
    \centering
    \begin{adjustbox}{minipage=\linewidth,scale=1}
        \centering
        \begin{subfigure}{0.49\textwidth}
            \resizebox{\textwidth}{!}{
                \includegraphics{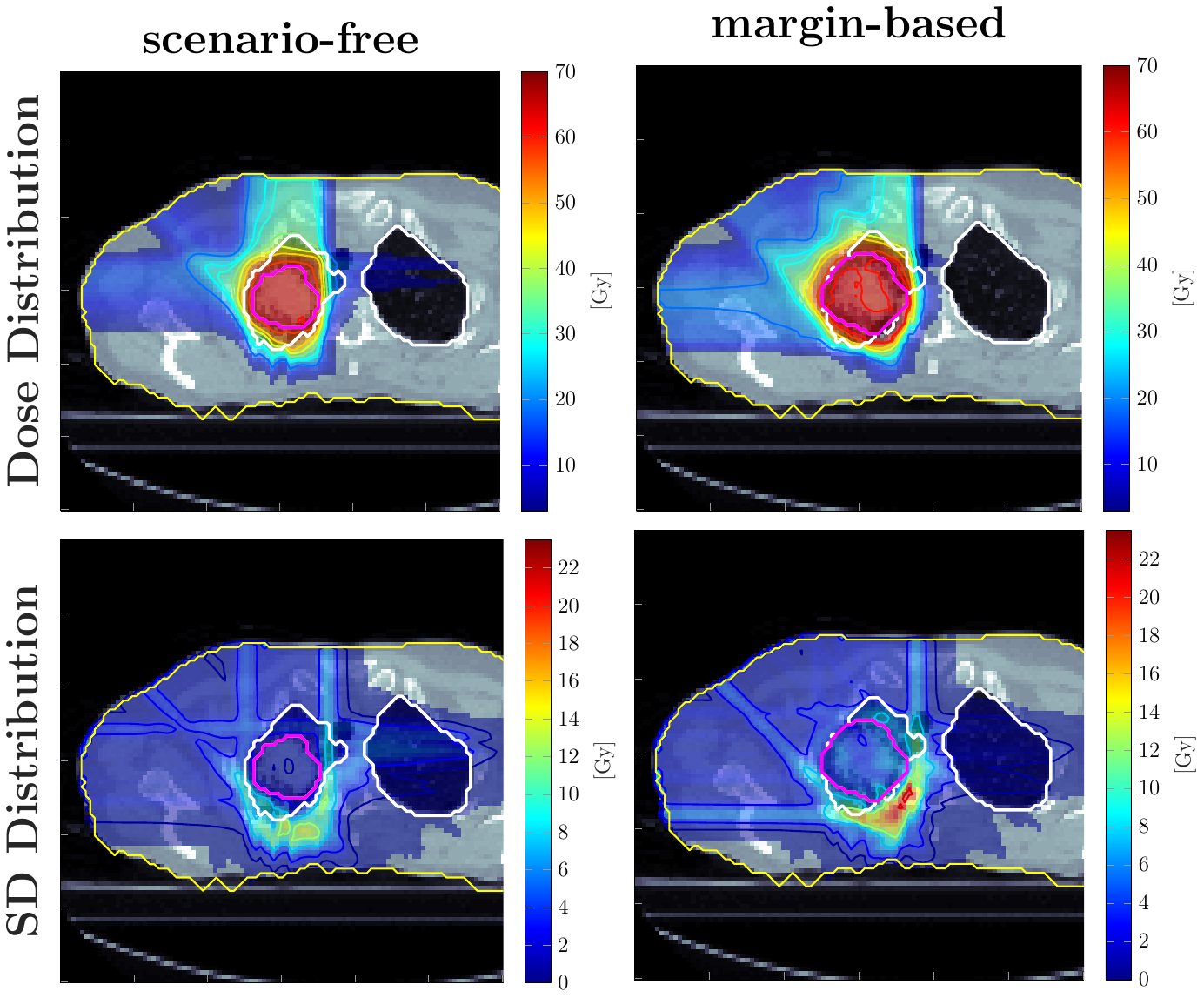}
              }
            \caption{Expected dose and SD distributions}
            \label{fig:P2-protons-panel_a}
        \end{subfigure}
        \hfill
        \begin{subfigure}{0.49\textwidth}
            \resizebox{\textwidth}{!}{
                \includegraphics{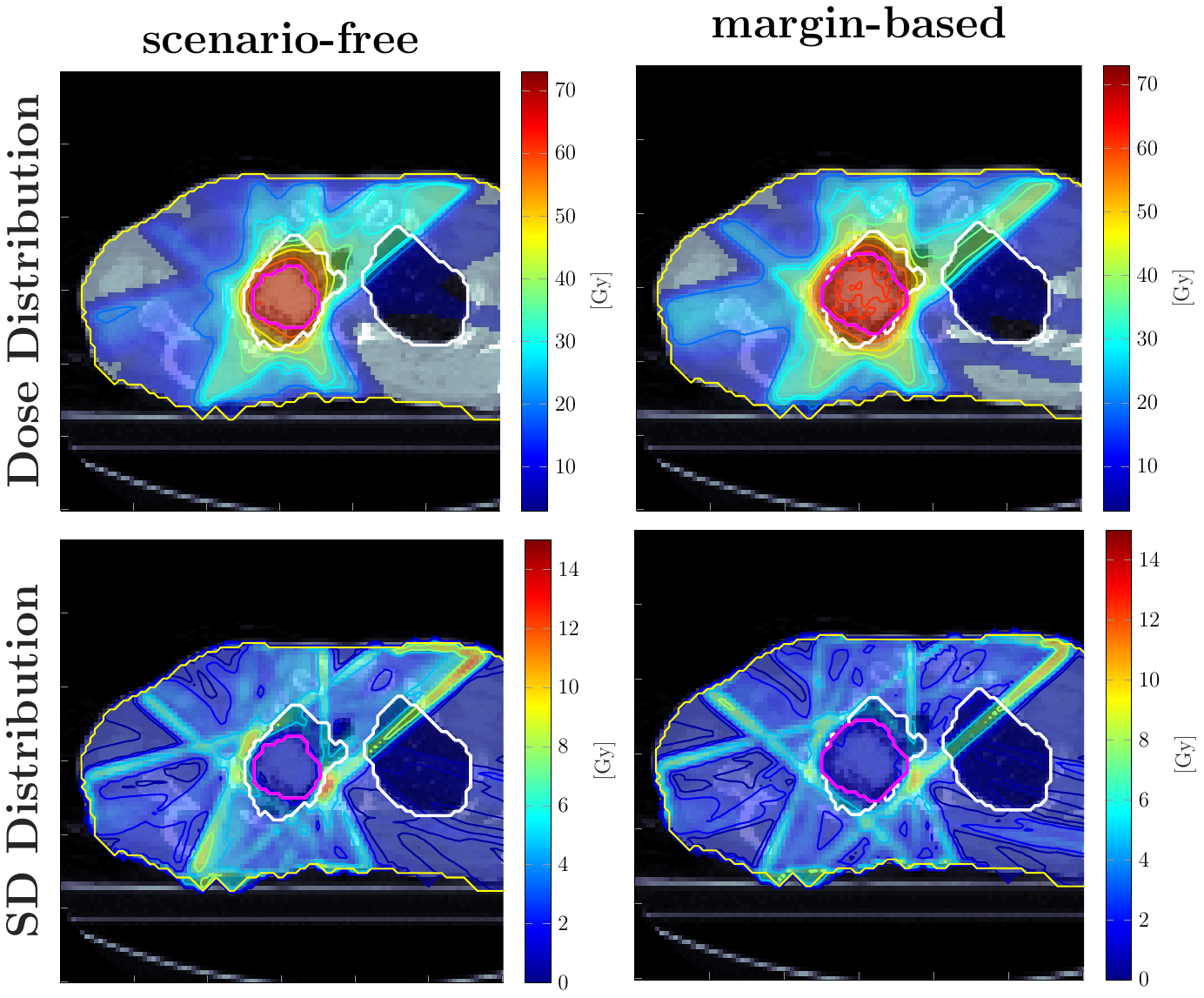}
            }
        \caption{Expected dose and SD distributions}
        \label{fig:P2-photons-panel_a}
        \end{subfigure}
    \end{adjustbox}
    \caption{Expected dose (top) and SD (bottom) distributions obtained with the scenario-free and the margin-based algorithms for the proton (a) and photon (b) irradiation plan. The contoured structures correspond to the lungs (white) and the target structure (purple). The target structure coincides with the CTV for the robust optimization, and with the PTV for the margin-based optimization.}
    \label{fig:P2-distributions}
\end{figure}

\begin{figure}[]
    \centering
    \begin{adjustbox}{minipage=\linewidth,scale=1}
        \begin{subfigure}[b]{0.49\textwidth}
            \resizebox{\textwidth}{!}{
                \includegraphics{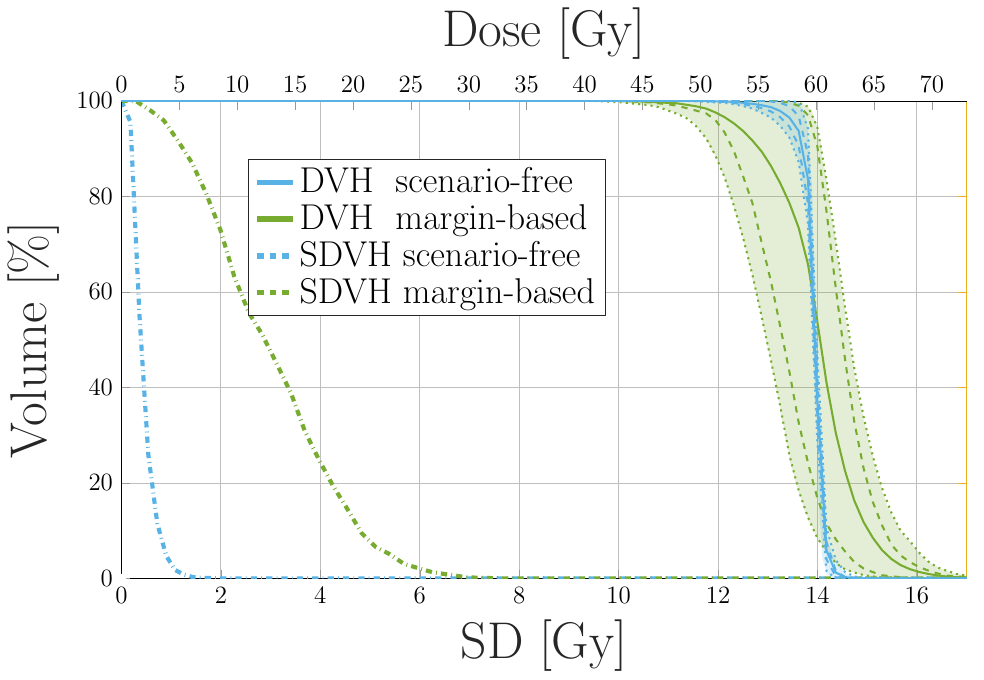}
            }
        \caption{Proton Target DVHs and SDVHs}
        \end{subfigure}
        \hfill
        \begin{subfigure}[b]{0.49\textwidth}
            \resizebox{\textwidth}{!}{
                \includegraphics{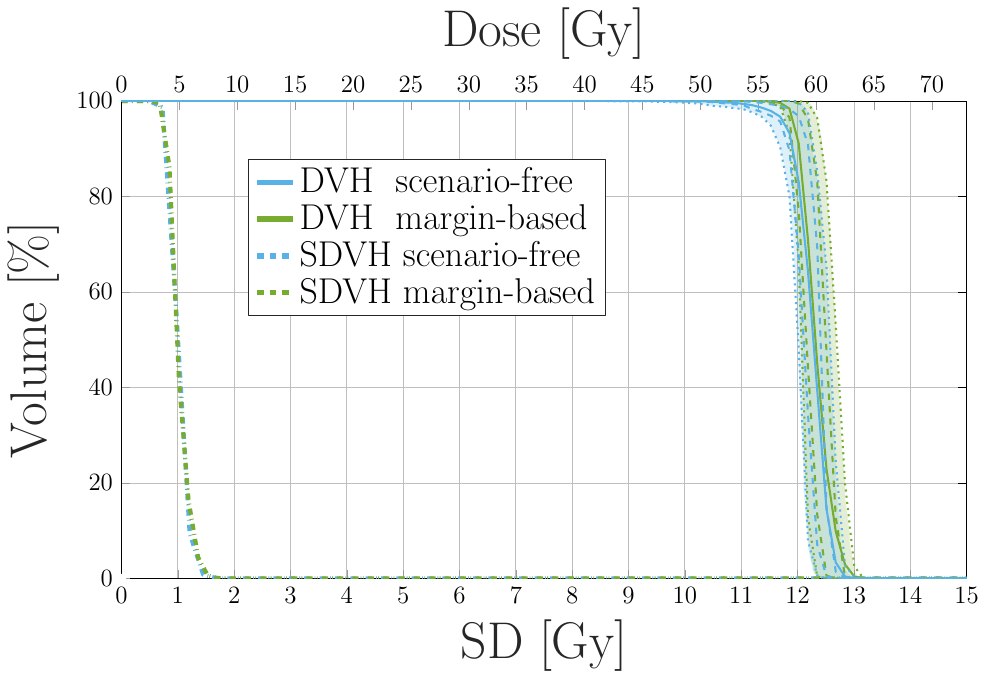}
            }
            \caption{Photons Target DVHs and SDVHs}
        \end{subfigure}
    
        \begin{subfigure}[b]{0.49\textwidth}
                \resizebox{\textwidth}{!}{
                    \includegraphics{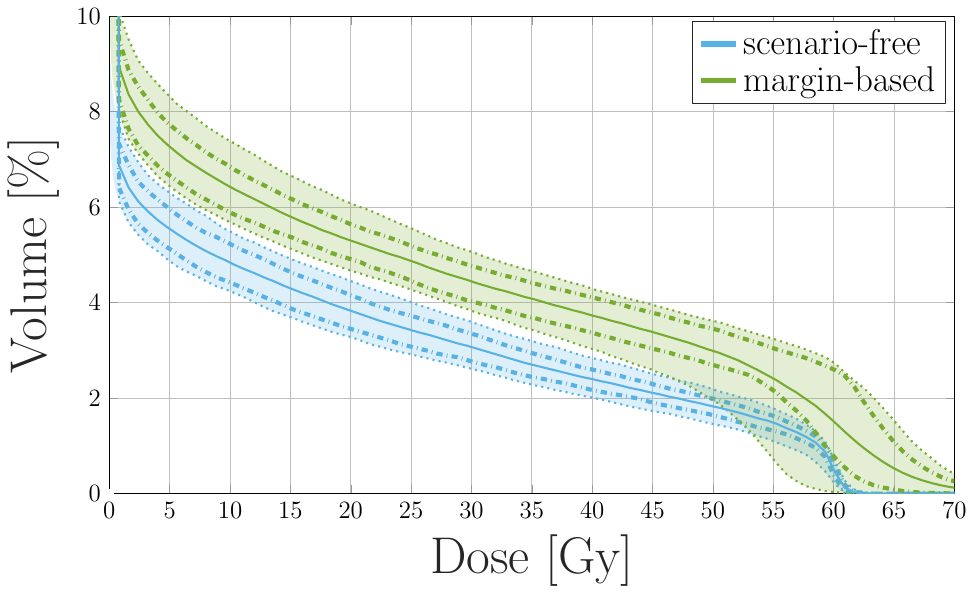}
                }
            \caption{Proton Lung DVHs}
            \label{fig:P2_protons_DVH_Lung}
        \end{subfigure}
        \hfill
        \begin{subfigure}[b]{0.49\textwidth}
            \resizebox{\textwidth}{!}{
                \includegraphics{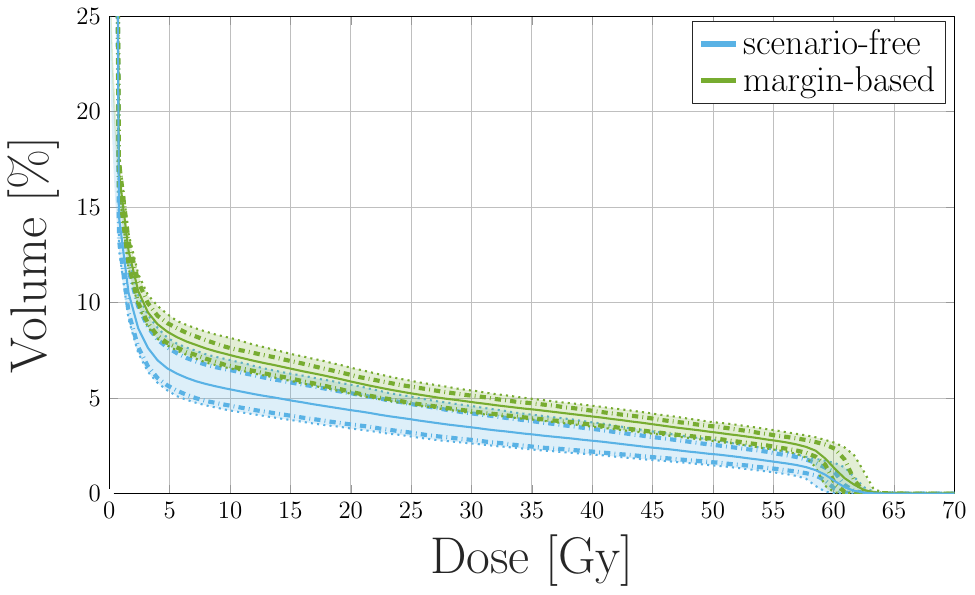}
            }
            \caption{Photons Lung DVHs}
            \label{fig:P2_photons_DVHs_d}
        \end{subfigure}

        \begin{subfigure}[b]{0.49\textwidth}
                \resizebox{\textwidth}{!}{
                    \includegraphics{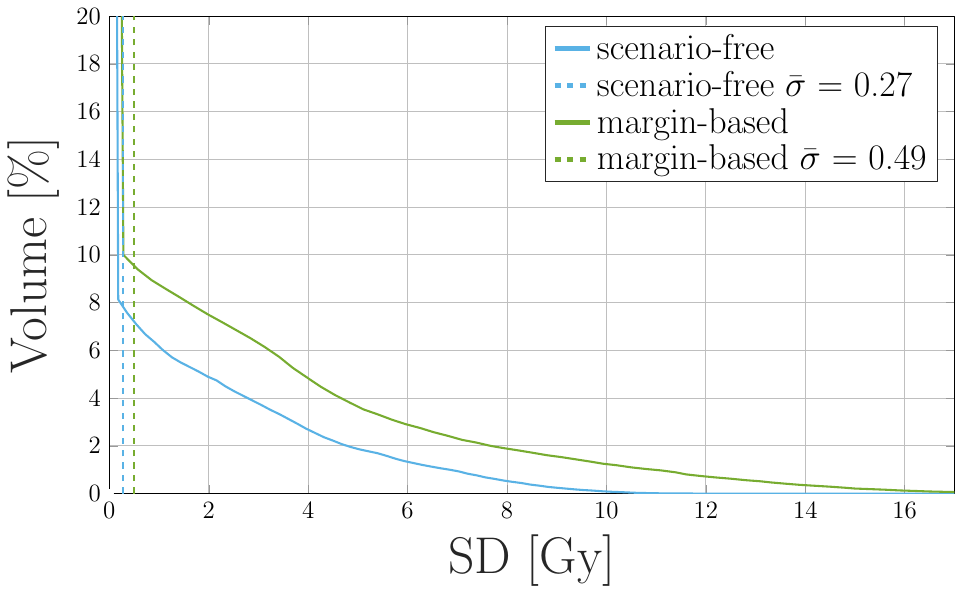}
                }
            \caption{Protons Lung SDVHs}
            %\label{fig:P1_proton_DVHs_e}
        \end{subfigure}
        \hfill
        \begin{subfigure}[b]{0.49\textwidth}
            \resizebox{\textwidth}{!}{
                \includegraphics{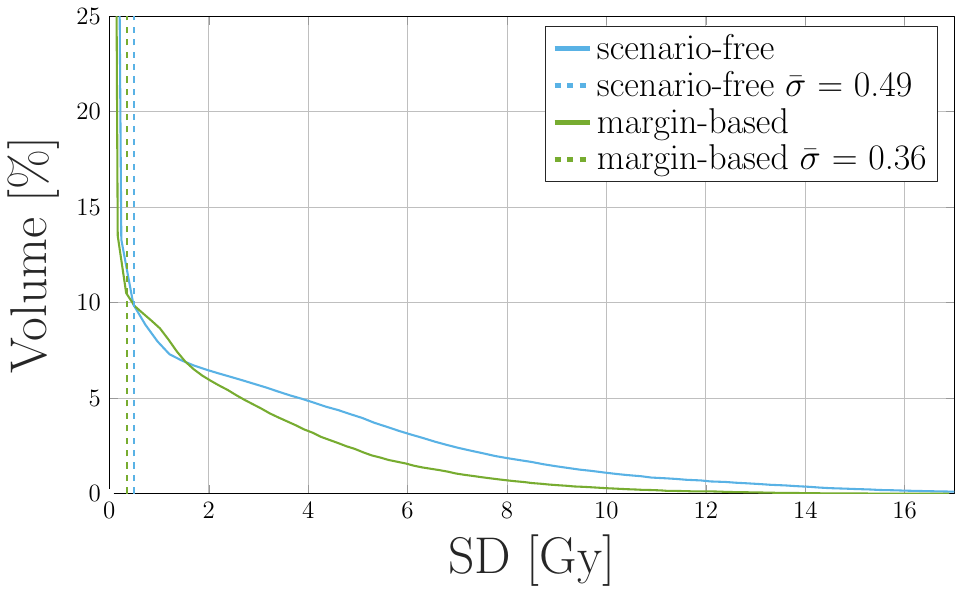}
            }
            \caption{Photons Lung SDVHs}
            %\label{fig:P1_proton_DVHs_f}
        \end{subfigure}
    \end{adjustbox}
    \caption{Dosimetric and robustness analysis performed for the photon (left) and proton (right) plans optimized with the scenarios-free and margin-based methods.
    For the reported DVHs, the solid line represents the DVH computed for the expected dose distribution while the dashed and dotted lines correspond to the \num{25}-\num{75} and the \num{5}-\num{95} percentiles of single-scenario DVHs distributions. The dotted vertical lines in the SDVH plot represent the mean standard deviation, i.e., the average value of the SD distribution.}
   \label{fig:P2_DVHs}
\end{figure}

\Cref{fig:P2-distributions} reports expected dose and SD distributions obtained with for the proton and photon plans. Additional dosimetric and robustness analysis was performed and reported in \cref{fig:P2_DVHs} for the same plans. For the proton case, the scenario-free approach was capable of delivering a more conformal dose distribution to the CTV. The SDVH curve for the target structure, as well as the dose delivered to the lung, were also lower with respect to the margin-based approach.

For the photon case the benefit in terms of target coverage and robustness of applying robust optimization is less evident when compared to the analogous proton plan. Both algorithms are indeed capable of achieving sufficient target coverage and the SDVH curves for the target structure are comparable. The scenario-free approach was again capable of reducing the overall dose delivered to the lung, as highlighted by the DVH distribution in \cref{fig:P2_protons_DVH_Lung,fig:P2_photons_DVHs_d}.

\begin{figure}[hbtp]
    \begin{subfigure}{0.5\textwidth}
        \resizebox{\textwidth}{!}{
            \includegraphics{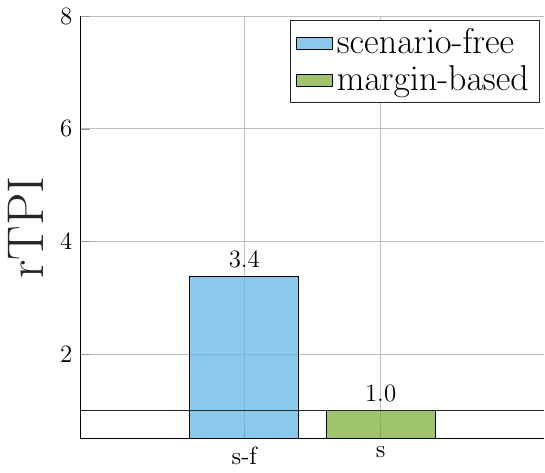}
        }
        \caption{Proton plan rTPI}
    \end{subfigure}
    \begin{subfigure}{0.5\textwidth}
        \resizebox{\textwidth}{!}{
            \includegraphics{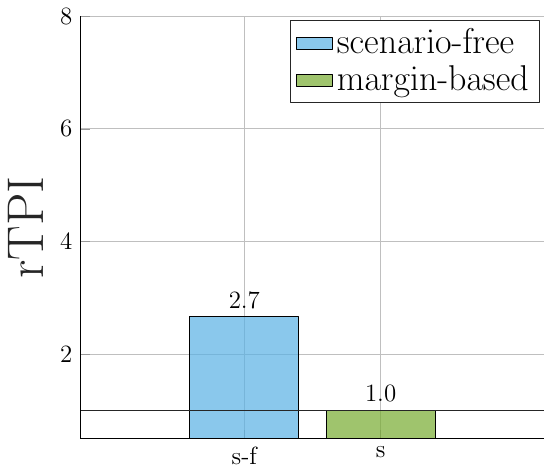}
        }
        \caption{Photon plan rTPI}
    \end{subfigure}
     \caption{rTPI for the proton (left) and photon (right) lung cancer patient plan. The abscissa labels refer respectively to the scenario-free (s-f) and (m-b) approaches.}
    \label{fig:P2_times}
\end{figure}

\Cref{fig:P2_times} reposrts the measured rTPIs for both radiation modalities. The observed computational times are compatible with the results obtained for the previous patient case and highlight how the scenario-free appraoch allows to gain robustness goals with minimial time overhead.

% Discussion
\section{Discussion}
The present work demonstrated the applicability and potentiality of a scenario-free approach to robust optimization for IMPT and IMRT. The scenario-free approach uses comparably low-dimensional, pre-computed quantities, that is, expected dose influence and total variance influence. During robust plan optimization, scenarios are thus not needed. While, in principle, a fully scenario-free pipeline can be constructed using, for example, analytical probabilistic modeling \cite{Bangert2013, Wahl2017, Wieser2017, Wahl2018a}, the scenario-free approach generalizes to precomputation from conventional error sampling during dose calculation, as used in this work, and is thus compatible with the current state-of-the-art in uncertainty modeling for radiotherapy.

\subsubsection{Validation}
Validation of the scenario-free algorithm was conducted exploiting the box-shaped phantom for both 3D and 4D optimization.
Initial comparison between plans optimized with pure least-square cost-functions proved the stability of the algorithm behavior. The comparison reported in \cref{fig:box_dose_dist} and \cref{fig:box_target_DVH_SDVH} highlights the compatibility of the scenario-free and stochastic algorithms when equivalency of the two approaches is explicit.

Beyond this equivalence, the scenario-free algorithm facilitates decoupling robustness by providing individual objective functions for variance (minimization) and expected dose at minimal computational effort. In the context of multi-criteria decision making, this separation makes robustness -- in the sense of variance -- an additional decision criterion in the planning process. The specific objective functions compete with the other conventional objectives and constraints, during optimization, and different trade-offs between nominal dose and robustness may be explored.

The addition of objectives different to the basic squared-deviation formulation to the global cost-function drops the equivalence between the scenario-free and stochastic approach. This can be understood comparing the distributions and analysis reported in \cref{fig:distributions_sf_sf_stoch_rnd_wc,fig:wc_5_to_100_panel} for the two approaches. Moving from a pure least-square optimization configuration to non-zero threshold squared-overdosing increases the feasibility of the dosimetric objective for the OARs, while keeping the variance reduction objective unbiased. This way the variance reduction objective gains an effective relative importance over the dosimetric objectives and results thus is a stronger focus of the optimization on the uncertainty minimization.

While this does not have consequences concerning overall convergence or convexity compared to the stochastic approach, it may distort the stochastical meaning. The approach no longer optimizes the expected value of the objective, but separately minimizes variance next to dose and expected dose objectives. The incontinent loss of intuitive interpretation for the stochastic approach as an expected value optimization is compensated by the possibility to directly quantify the uncertainty impact in the form of mean variance, thus relying the effective application of the scenario-free algorithm upon the understanding and characterization of the variance reduction objectives' impact.

Being the optimization time and memory footprint for the scenario-free approach independent on the number of scenarios, its application to contexts involving a large dimensionality of the uncertainty model is particularly beneficial. 4D robust optimization is, in this regard an highly demanding approach that combines setup and range uncertainties with the multiple CT representations of the patient.

The scenario-free approach allows, on one hand, for the inclusion of a large scenario sample size in the calculation of the $\boldsymbol{\Omega}$ and $\mathbb{E}[\boldsymbol{D}]$ matrices, improving the statistical representation of the uncertainty model and the effectiveness of the robust optimization. On the other hand, it also offers the possibility to focus on different sources of uncertainty by appropriately pooling together the error scenarios. In 4D optimization, this expresses the possibility to define total and phase-specific probabilistic quantities and thus address the impact of specific deviations. The analysis reported in \cref{fig:4D-Box-phase-vs-all} showcases this possibility and highlights how, for this particular instance, no significant dosimetric benefit is achieved in limiting the pooling of scenarios to the specific CT-phases.

As a general remark, the $\boldsymbol{\Omega}$ matrix encodes correlations between beamlets. If the beamlet representation has a direct physical interpretation, as is the case for pencil beam scanning in IMRT, and the scanning pattern is known, the 4D CT-phases sampling the patient motion can be appropriately coupled with subsets of beamlets to model interplay effects \cite{Stammer2023}. By pooling together phase-specific scenarios for the calculation of the $\boldsymbol{\Omega}$ matrices, the impact of interplay effects would also be captured by such construct and minimized.

\subsubsection{4D Patient Cases}
After phantom validation, the approach was tested on two realistic lung patients. The analysis performed resulted in comparable target coverage and dose delivered to the OARs between the two robust optimization approaches. For both the scenario-free approach and the stochastic reference implementation, target coverage and plan robustness was improved over the margin-based approach. This is highlighted by the DVH and SDVH lines reported in \cref{fig:P1_proton_DVHs}, and by the box-plot metrics in \cref{fig:P1_BoxPlots}. Even in this case, the scenario-free approach was ultimately able to achieve plan quality requirements as expected from a robustly optimized plan, while bearing only slightly higher computation cost during optimization compared to the nominal margin-based optimization.

Compared to the simplified phantom geometry, dimensionality of the dose-influence matrix increased heavily due to the larger number of individual spots (due to higher number and dimensions of the fields). Calculation of multiple sparse matrix-vector products becomes thus more limiting, in storage as well as in computation time. While the limit to which these data structures and computations are feasible is dependent on the specific architecture and implementation, the linear storage and runtime complexity will eventually break the approach and increase discrepancy between the optimization times of the stochastic and scenario-free approaches. The application of the scenario-free approach is indeed unaffected by the number of scenarios and thus by the dimensionality of the influence matrices, apart from precomputation if scenarios are used for construction.

Consequently, the number of scenarios included in the optimization and analysis of this patient case was limited by the applicability of the stochastic robust optimization. For the specific machine used and setup, \num{30} was the maximum number of scenarios that could be included with the specific system and reference implementation without reaching physical memory limitations.

The second patient case included instead \num{100} total scenarios. The stochastic robust optimization algorithm could not be applied with the given specifications due to the limitations in available memory. This case highlights how the scenario-free optimization algorithm allows to achieve robustness goals in setups where the traditional algorithms would fail or became unfeasible.

In this case, the DVH distributions for the target in the photon plan did not highlight significant differences between the two optimization algorithms. The reason for this behavior should be found again in the static dose cloud approximation. This still holds to some extent for the photon plan, and the PTV margin is sufficient to preserve target coverage. At the same time, the DVH analysis for the Lung highlighted how the scenario-free approach is capable of reducing the dose delivered to this OAR.

The scenario-free proton plan on the contrary, achieved significantly better target coverage and reduction of the SD distribution both within the target and the main OARs. This behavior is expected given the higher sensitivity of the proton dose distribution to setup and range uncertainty.

Both the photon and proton robust optimization for this patient case optimized a robust dose distribution including a large number of scenarios within an optimization time comparable to a nominal optimization.

\subsection{Summary \& Outlook}
Overall, this work showed how the scenario-free approach can achieve the desired plan quality and robustness while significantly reducing the required optimization time and memory.

In a realistic 4D optimization setup, the number of scenarios can easily rise due to the multiplicity of CT-phases and error scenarios. Optimization of a robust dose distribution in such cases is necessarily burdened by the computational cost.
By avoiding the storage of multiple dose-influence matrices and the estimation of as many dose distributions at each iteration step, the scenario-free optimization performance is unaffected by the number of CT-phases or error scenarios included. 

The absolute optimization time for both algorithms is determined by the specific hardware configuration used and by the algorithm implementation. Compatible optimization conditions were applied to ensure consistency of the measured relative computation times, i.e. the same machine was used to perform all calculations. Additionally, both robust approaches were developed and applied within the multi-purpose open source treatment planning toolkit \emph{matRad} and no method-specific code optimization was performed to enhance nor hinder the performances of a specific algorithm.

This also suggests the possibility for further improvement of both algorithms performances if dedicated architecture-specific algorithm implementations were to be applied. The intrinsic symmetry of the $\boldsymbol{\Omega}$ matrix could for example be exploited to reduce the numerical workload in the estimation of the variance terms for the scenario-free approach, which could potentially benefit from a dedicated GPU implementation.
\\

Both the considered robust optimization approaches rely on the calculation of multiple scenario-specific dose-influence matrices prior to optimization. While the stochastic algorithm makes explicit use of such matrices during optimization, the scenario-free approach still requires an intermediate step for the accumulation of the probabilistic quantities $\boldsymbol{\Omega}$ and $\mathbb{E}[\boldsymbol{D}]$. The time overhead required to perform such accumulation varies according to the dimensionality and number of the dose-influence matrices involved and has to be accounted for while considering the complete plan development workflow.

The largest benefit obtained with the scenario-free approach in optimization time reduction and workflow streamline is therefore achieved when multiple plan optimizations can be performed with the same precalculated probabilistic quantities. This is the case of traditional trial and error penalized weighted-sum approaches, were the relative importance of optimization objectives and constraints is manually tuned to meet the plan prescriptions. More sophisticated interactive and automated multi-criteria optimization (MCO) planning procedures such as Pareto front navigation and Lexicographic approaches were developed to overcome the limitations of such traditional approaches. These techniques intrinsically require multiple plans to be subsequently optimized and would thus benefit from any reduction of the optimization time required.

Additionally, previous work reported the applicability of robust optimization to MCO techniques \cite{Chen2012}. With its capability to independently quantify and address robustness in terms of a variance minimizing objective, the scenario-free approach offers the possibility to introduce robustness as an additional tradable dimensionality to the Pareto front definition of a MCO problem. Consistently, variance reduction could be applied as an intermediate or final optimization step in an otherwise nominal prioritized lexicographic approach as well.

Addition of specific structure robustifying objectives on top of a nominal optimization could find a potential application in an adaptive planning workflow as well. The daily adapted plan could be obtained with a fast nominal optimization and the voxel-independent variance reduction chained subsequently. Calculation of the probabilistic quantities could then be based on multiple modeled sources of uncertainties together with the patient specific deviations and inter-fractional anatomical changes observed through the course of treatment. This would potentially allow achievement a stable optimized dose distribution with minimal additional computational effort.

% Conclusion
\section{Conclusions}
This work demonstrates the capabilities and potentiality of a scenario-free robust optimization method. The algorithm was applied to several configurations, involving both artificial and clinical 3D and 4D robust optimization setups. The conducted analysis highlighted how this approach allows computationally and memory efficient optimization of robust treatment plans for IMRT and IMPT. The quality and robustness of such plans is at least comparable with those of plans obtained with traditional optimization strategies, while the optimization time observed for the scenario-free approach is severely reduced with respect to the traditional robust optimization and indeed comparable to the nominal, non-robust approach.

The scenario-free approach allows to include a large number of error scenarios in the development of the treatment plan, thus to achieve solid robustness goals, while keeping the memory requirements and the optimization time within feasible ranges.

Finally, the algorithm facilitates selective reduction of uncertainty within given structures by introducing specific variance reduction cost-functions. These features make the scenario-free algorithm suitable for highly computationally demanding optimization configurations such as 4D robust optimization.

Beyond the applications shown in this analysis, MCO interactive and automated planning strategies could significantly benefit from the runtime reduction and the variance reduction flexibility offered by the scenario-free approach. Additional flexibility is gained with different accumulation strategies for the probabilistic quantities, which could in principle be designed to include beam delivery uncertainties such as interplay effects as well.

The efficiency achievable with this approach could open the path to the introduction of robust optimization to highly runtime-constraints optimization scenarios such as an adaptive radiotherapy workflow.

\section{Acknowledgement}
Funding by the Deutsche Forschungsgemeinschaft (DFG, German Research Foundation), Projects No. 443188743 and 457509854, is acknowledged.

\section{Conflict of interest}
The authors declare that they have no known competing financial interests or personal relationships that could have appeared to influence the work reported in this paper.

\printbibliography

\end{document}